\UseRawInputEncoding
\documentclass[review]{elsarticle}

\usepackage{lineno,hyperref}
\usepackage{float}
\usepackage{amsmath,amssymb,amsfonts}
\usepackage{algorithmic}
\usepackage{lipsum}
\usepackage{graphicx}
\usepackage{textcomp}
\usepackage{sidecap}
\usepackage{caption}
\usepackage{subcaption}
\usepackage{xfrac}
\usepackage{textcomp}
\usepackage{hyperref}
\usepackage[dvipsnames]{xcolor}
\nolinenumbers

\usepackage{verbatim}
\usepackage{setspace}
\usepackage[belowskip=2pt,aboveskip=10pt]{caption}

\makeatletter
\let\old@rule\@rule
\def\@rule[#1]#2#3{\textcolor{rulecolor}{\old@rule[#1]{#2}{#3}}}
\makeatother

\definecolor{rulecolor}{named}{White}

\journal{xxx}

\biboptions{authoryear}

\begin{document}

\begin{frontmatter}

\title{Deconvolution of the Functional Ultrasound Response in the Mouse Visual Pathway Using Block-Term Decomposition}

% %% Group authors per affiliation:
% \address{Radarweg 29, Amsterdam}
% \fntext[myfootnote]{Since 1880.}


\author[delftaddress]{Aybuke Erol\corref{mycorrespondingauthor}}
\cortext[mycorrespondingauthor]{Corresponding author.}
\ead{a.erol@tudelft.nl}
\author[cubeaddress]{Chagajeg Soloukey}
\author[cubeaddress]{Bastian Generowicz}
\author[cubeaddress]{Nikki van Dorp}
\author[cubeaddress]{Sebastiaan Koekkoek}
\author[cubeaddress]{Pieter Kruizinga}
\author[delftaddress]{Borbala Hunyadi}

\address[delftaddress]{Circuits and Systems (CAS), Department of Microelectronics, Delft University of Technology, Mekelweg 5, 2628 CD Delft, The Netherlands}
\address[cubeaddress]{Center for Ultrasound and Brain imaging at Erasmus MC (CUBE), Department of Neuroscience, Erasmus Medical Center, Doctor Molewaterplein 40, 3015 GD Rotterdam, The Netherlands}

\begin{abstract}
Functional ultrasound (fUS) indirectly measures brain activity by recording changes in cerebral blood volume and flow in response to neural activation. Conventional approaches model such functional neuroimaging data as the convolution between an impulse response, known as the hemodynamic response function (HRF), and a binarized representation of the input (i.e., source) signal based on the stimulus onsets, the so-called experimental paradigm (EP). However, the EP may not be enough to characterize the whole complexity of the underlying source signals that evoke the hemodynamic changes, such as in the case of spontaneous resting state activity. Furthermore, the HRF varies across brain areas and stimuli. To achieve an adaptable framework that can capture such dynamics and unknowns of the brain function, we propose a deconvolution method for multivariate fUS time-series that reveals both the region-specific HRFs, and the source signals that induce the hemodynamic responses in the studied regions. We start by modeling the fUS time-series as convolutive mixtures and use a tensor-based approach for deconvolution based on two assumptions: (1) HRFs are parametrizable, and (2) source signals are uncorrelated. We test our approach on fUS data acquired during a visual experiment on a mouse subject, focusing on three regions within the mouse brain's colliculo-cortical, image-forming pathway: the lateral geniculate nucleus, superior colliculus and visual cortex. The estimated HRFs in each region are in agreement with prior works, whereas the estimated source signal is observed to closely follow the EP. Yet, we note a few deviations from the EP in the estimated source signal that most likely arise due to the trial-by-trial variability of the neural response across different repetitions of the stimulus observed in the selected regions.
\end{abstract}

\begin{keyword}
hemodynamic response, HRF, deconvolution, functional ultrasound, tensor decomposition, mouse, visual perception
\end{keyword}

\end{frontmatter}

\section{Introduction}

Functional ultrasound (fUS) is a neuroimaging technique that indirectly measures brain activity by detecting changes in cerebral blood flow (CBF) and volume (CBV) \citep{fusrbc}. The fUS signal is related to brain activity through a process known as neurovascular coupling (NVC). When a brain region becomes active, it calls for an additional supply of oxygen-rich blood, which creates a hemodynamic response (HR), i.e., an increase of blood flow to that region. NVC describes this interaction between local neural activity and blood flow \citep{b2}. Functional ultrasound is able to measure the HR because of its sensitivity to fluctuations in blood flow and volume \citep{b1}. In the past decade, fUS has been successfully applied in a variety of animal and clinical studies, showing the technique's potential for detection of sensory stimuli, as well as complex brain states and behavior \citep{fus_npixels}. These include studies on small rodents \citep{param1,setup,cube1}, birds \citep{rau} and humans \citep{sadaf,humanfus,humanfus2}.

Understanding the HR has been an important challenge not only for fUS \citep{b3}, but also for several other established functional neuroimaging modalities, including functional magnetic resonance imaging (fMRI) \citep{b4} and functional near-infrared spectroscopy (fNIRS) \citep{b5}. The HR can be characterized by a function representing the impulse response of the neurovascular system, known as the hemodynamic response function (HRF) \citep{b6}. To form a model for the HR, the HRF gets convolved with an input signal representing the experimental paradigm (EP), which is expressed as a binary vector that shows the on- and off- times of a given stimulus. However, not all brain activity can be explained via such predefined and external stimuli \citep{b13}. Indeed, even when no stimulus is presented, there can still be spontaneous, non-random activity in the brain, reported to be as large as the activity evoked by intentional stimulation \citep{Gilbert}. Therefore, the input signals that trigger brain activity should be generalized beyond merely the EP. This issue has been addressed by \citep{b13,actinduc2,actinduc}, where the authors have defined the term \emph{activity-inducing} signal, which, as the name suggests, comprises any input signal that induces hemodynamic activity. We will refer to activity-inducing signals as \emph{source signals} in the rest of this paper, which steers the reader to broader terminology not only used in biomedical signal processing, but also in acoustics and telecommunications \citep{sources}, and emphasizes that recorded output data are \emph{sourced} by such signals. 

An accurate estimation of the HRF is crucial to correctly interpret both the hemodynamic activity itself and the underlying source signals. Furthermore, the HRF has shown potential as a biomarker for pathological brain functioning, examples of which include obsessive-compulsive disorder \citep{hrfocd}, mild traumatic brain injury \citep{hrfinjury}, Alzheimer's disease \citep{hrfdementia}, epilepsy \citep{eegfmri} and severe psychosocial stress \citep{hrfstress}. While HRFs can as well be defined in nonlinear and dynamic frameworks with the help of Volterra kernels \citep{volterra}, linear models have particularly gained popularity due to the combination of their remarkable performance and simplicity. Several approaches have been proposed in the literature which employ linear modelling for estimating the HRF. The strictest approach assumes a constant a priori shape of the HRF, i.e. a mathematical function with fixed parameters, and is only concerned with finding its scaling (the activation level). The shape used in this approach is usually given by the canonical HRF model \citep{b7}. As such, this approach does not incorporate HRF variability, yet the HRF is known to change significantly across subjects, brain regions and triggering events \citep{b8, hrfchange1, hrfchange2}. A second approach is to estimate the parameters of the chosen shape function, which provides a more flexible and unbiased solution  \citep{b3}. Alternatively, HRF estimation can be reformulated as a regression problem by expressing the HRF as a linear combination of several basis functions (which are often chosen to be the canonical HRF and its derivatives). This approach is known as the general linear model (GLM) \citep{b9}. Finally, it is also possible to apply no shape constraints on the HRF, and predict the value of the HRF distinctly at each time point. This approach suffers from high computational complexity and can result in arbitrary or physiologically meaningless forms \citep{b10}.

Note that the majority of studies which tackle HRF estimation presume that the source signal is known and equal to the EP, leaving only one unknown in the convolution: the HRF \citep{neuralknown}. However, as mentioned earlier, a functional brain response can be triggered by more sources than the EP alone. These sources can be extrinsic, i.e., related to environmental events, such as unintended background stimulation or noise artefacts. They might also be intrinsic sources, which can emerge spontaneously during rest \citep{rest}. Under such complex and multi-causal circumstances, recovering the rather 'hidden' source signal(s) can be of interest. Moreover, even the EP itself can be much more complex than what a simple binary pattern allows for. Indeed, the hemodynamic response to, for instance, a visual stimulus, can vary greatly depending on its parameters, such as its contrast \citep{param1} or frequency \citep{param2}. %In terms of duration, for example, the shape of an HRF evoked by a long-lasting boxcar stimulus has a sustained peak, which differs from the HRF of a short, single impulse stimulus \citep{b27}. 
In contrast to the aforementioned methods, where the goal was to estimate HRFs from a known source signal, there have also been attempts to predict the sources by assuming a known and fixed HRF \citep{b12} \citep{b13}. However, these methods fall short of depicting the HRF variability.

To sum up, neither the sources nor the HRF are straightforward to model, and as such, when either is assumed to be fixed, it can easily lead to misspecification of the other. Therefore, we consider the problem of jointly estimating the source signals and HRFs from multivariate fUS time-series. This problem has been addressed by \citep{b14}, \citep{b15} and \citep{b16}. In \citep{b14}, it is assumed that the source signal (here considered as the neural activity) lies in a high frequency band compared to the HRF, and can thus be recovered using homomorphic filtering. On the other hand, \citep{b15} first estimates a spike-like source signal by thresholding the fMRI data and selecting the time points where the response begins, and subsequently fits a GLM using the estimated source signal to determine the HRF. Both of the mentioned techniques share the limitation of being univariate methods: although they analyze multiple regions and/or subjects, the analysis is performed separately on each time series, thereby ignoring any mutual information shared amongst biologically relevant ROIs. 

Recently, a multivariate deconvolution of fMRI time series has been proposed in \citep{b16}. The authors proposed an fMRI signal model, where neural activation is represented as a low-rank matrix - constructed by a certain (low) number of temporal activation patterns and corresponding spatial maps encoding functional networks - and the neural activation is linked with the observed fMRI signals via region-specific HRFs. The main advantage of this approach is that it allows whole-brain estimation of HRF and neural activation. However, all HRFs are defined via the dilation of a presumed shape, which may not be enough to capture all possible variations of the HRF, as the width and peak latency of the HRF are coupled into a single parameter. Moreover, the estimated HRFs are region-specific, but not activation-specific. Therefore, the model cannot account for variations in the HRF due to varying stimulus properties. Yet, the length and intensity of stimuli appear to have a significant effect on HRF shape even within the same region, as observed in recent fast fMRI studies \citep{stimhrf}. 

In order to account for the possible variations of the HRF for both different sources and regions, we model the fUS signal in the framework of convolutive mixtures, where multiple input signals (sources) are related to multiple observations (measurements from a brain region) via convolutive mixing filters. In the context of fUS, the convolutive mixing filters stand for the HRFs, which are unique for each possible combination of sources and regions, allowing variability across different brain areas and triggering events. In order to improve identifiability, we make certain assumptions, namely that the shape of the HRFs can be parametrized and that the source signals are uncorrelated. Considering the flexibility of tensor-based formulations for the purpose of representing such structures and constraints that exist in different modes or factors of data \citep{b19}, we solve the deconvolution by applying block-term decomposition (BTD) on the tensor of lagged measurement autocorrelation matrices.

While in our previous work \citep{b20} we had considered a similar BTD-based deconvolution, this paper presents several novel contributions. First, we improve the robustness of the algorithm via additional constraints and a more sophisticated selection procedure for the final solution from multiple optimization runs. We also present a more detailed simulation study considering a large range of possible HRF shapes. Finally, instead of applying deconvolution on a few single pixel time-series, we now focus on fUS responses of entire ROIs, as determined by spatial independent component analysis (ICA). The selected ROIs represent three crucial anatomical structures within the mouse brain's colliculo-cortical, image-forming visual pathway: the lateral geniculate nucleus (LGN), the superior colliculus (SC) and the primary visual cortex (V1). These regions are vision-involved anatomical structures of importance \citep{huberman, seabrook}, which can be captured together well in a minimal number of coronal and sagittal slices \citep{bregma}, and have proven to consistently yield clear responses using fUS imaging \citep{param1,mace_visual}. %Together, these three anatomical regions form the strongest two pathways linking the eye to the brain. 

The vast majority of information about visual stimuli are conveyed via the retinal ganglion cells (RGCs) to the downstream subcortical targets LGN and SC, before being relayed to V1.% A higher order region such as V1 can also influence subcortical areas like SC [ref]. 
The LGN and SC are known to receive both similar and distinct visual input information from RGCs \citep{sclgn}. The asymmetry in information projected by the mouse retina to these two downstream targets is reflected in the output of these areas \citep{Ellis}. % For instance, some RGCs - those which transiently and selectively respond to small stimuli - only project to the SC.  Vascular responses of the LGN, for example, have been shown to differ depending on duration and frequency of stimulus input \citep{alpha}, orientation and direction of a moving stimulus \citep{gamma}, (lack of) contrast \citep{gamma, beta}, as well as the mouse’s state of wakefulness \citep{beta}. %The LGN is a laminated, retinotopically organized thalamic relay centre which is directly and reciprocally connected to V1.  Previous circuit analysis studies on the mouse SC, a laminated midbrain area where information from visual, auditory and somatosensory modalities are integrated into eye and head movements, reveal its key role in visual processing \citep{ItoSC2018}. Receiving projections from 85-90\% of RGCs \citep{Ellis}, mouse SC neurons are ‘feature detectors’, meaning that subsets of SC neurons will respond differently depending on the stimulus type that was presented within its receptive field. The SC is therefore thought to be useful for the swift visual detection of visual features that indicate potential threats (such as flashing, moving or looming spots \citep{Gale, ItoSC2017, Zhao, Wang, Inayat}). % The mouse visual cortex has been reported to include up to $16$ interconnected, retinotopically organized areas \citep{Zhuang}. Different areas of the visual cortex are preferentially activated by different spatiotemporal stimulus frequencies \citep{param1, Marshel, Andermann, Tohmi, param2, Roth}. Although the mouse LGN, SC and V1 have all been shown to modulate their response based on visual stimulus features, we are unsure how each regions' specific characteristics will reflect in terms of the fUS signal.
Our goal is to compare the hemodynamic activity in these regions by deconvolving the CBF/CBV changes recorded with fUS in response to visual stimulus.

% We have proposed convolutive mixtures modelling of fUS signals and the corresponding tensor solution in our previous work \citep{b20}, where we have worked on a few pixel time series. In this paper, we add a preparatory experiment to choose an optimal EP for the formulated blind deconvolution problem, improve on our simulation study, and extend our analysis to the whole brain. To be able to work on the whole brain, we start with a dimensionality reduction stage in space. We first apply ICA on the measurement matrix of all pixels, and determine the spatial signatures of interest. We later threshold these signatures to convert them into spatial masks, and average the pixel time series that are covered within these masks. This way, we aim to have a representative set of time series, calculated from groups of pixels that are both temporally and spatially coherent. In the end, we reduce both the dimensionality of the tensor of lagged measurement autocorrelations for BTD, and lower the amount of results to be interpreted to an admittedly smaller, but thereby more informative subset.

The rest of this paper is organized as follows. First, we describe our data model and the proposed tensor-based solution for deconvolution. Next, we describe the experimental setup and data acquisition steps used for fUS imaging of a mouse subject. This is followed by the deconvolution results, which are presented in two-folds: \emph{(i)} Numerical simulations, and \emph{(ii)} Results on real fUS data. Next, under discussion, we review the highlights of our modelling and results, and elaborate on the neuroscientific relevance of our findings. Finally, we state several future extensions and conclude our paper.

\section{Signal Model}

Naturally fUS images contain far more pixels than the number of anatomical or functional regions. We therefore expect
certain groups of pixels to show similar signal fluctuations, and we consider the fUS images as parcellated in space into several regions. Consequently, we represent the overall fUS data as an $M \times N$ matrix, where each of the $M$ rows contain the average pixel time-series within a region-of-interest (ROI), and $N$ is the number of time samples.

Assuming a single source signal, an individual ROI time-series $y(t)$ can be written as the convolution between the HRF $h(t)$ and the input source signal $s(t)$ as:
\begin{equation}
    \label{eq:singleconv}
        y(t) = \sum_{l=0}^L h(l)s(t-l)
\end{equation}
where $L+1$ gives the HRF filter length. 

However, a single ROI time-series may be affected by a number of ($R$) different source signals. Each source signal $s_r(t)$ may elicit a different HRF, $h_r(t)$. Therefore, the observed time-series is the summation of the effect of all underlying sources:
\begin{equation}
    \label{eq:ins_ica}
        y(t) = \sum_{r=1}^R \sum_{l=0}^L h_r(l)s_r(t-l).
\end{equation}

Finally, extending our model to multiple ROIs, where each ROI may have a different HRF, we arrive to the following multivariate convolutive mixture formulation:
\begin{equation}
    \label{eq:convolutive}
        y_m(t) = \sum_{r=1}^R \sum_{l=0}^L h_{mr}(l)s_r(t-l)
\end{equation}
where $h_{mr}(l)$ is the convolutive mixing filter, belonging to the ROI $m$ and source $r$ \citep{b21}.

In the context of fUS, the sources that lead to the time-series can be task-related ($T$), such as the EP, or artifact-related ($A$). The task-related sources are convolved with an HRF, whereas the artifact-related sources are directly additive on the measured time-series \citep{b22}. Yet, the strength of the effect that an artifact source exerts on a region should still depend on the artifact type and the brain region. To incorporate this in Eq. \ref{eq:convolutive}, each $h_{mr}(l)$ with $r \in A$ should correspond to a scaled (by $a_{mr}$) unit impulse function. Thus, we rewrite Eq. \ref{eq:convolutive} as:
\begin{align}
    \label{eq:convolutive2}
    y_m(t) &= \sum_{r\in T} \sum_{l=0}^L h_{mr}(l)s_r(t-l)+\sum_{r\in A} \sum_{l=0}^{L} a_{mr} \delta(l)s_r(t-l) \nonumber \\
    &= \sum_{r\in T} \sum_{l=0}^L h_{mr}(l)s_r(t-l)+\sum_{r\in A} a_{mr} s_r(t).
\end{align}

We aim at solving this deconvolution problem to recover the sources ($s_r,r\in T$) and HRFs ($h_{mr}, r\in T$) of interest separately at each ROI $m$. 

\section{Proposed Method} 

In this section, we will present the steps of the proposed tensor-based deconvolution method. We will first introduce how deconvolution of the observations modeled as in Eq. \ref{eq:convolutive2} can be expressed as a BTD. Due to the fact that this problem is highly non-convex, we will subsequently explain our approach to identifying a final solution for the decomposition. Finally, we will describe source signal estimation using the HRFs predicted by BTD.

\subsection{Formulating the Block-Term Decomposition}

We start by expressing the convolutive mixtures formulation in Eq. \ref{eq:convolutive} in matrix form as $\mathbf{Y}=\mathbf{H}\mathbf{S}$. The columns of $\mathbf{Y}$ and $\mathbf{S}$ are given by $\mathbf{y}(n)$, $n=1,\dots,N-L'$ and $\mathbf{s}(n)$, $n=1,\dots,N-(L+L')$, respectively. These column vectors are constructed as follows \citep{b23}:
\begin{align}
\begin{aligned}
    \label{eq:matrix_y_s}
    \mathbf{y}(n) &= [y_1(n),...,y_1(n-L'+1), \\ 
    & ...,y_M(n),...,y_M(n-L'+1)]^T\; \; \text{and}\\
    \mathbf{s}(n) &= [s_1(n),...,s_1(n-(L+L')+1), \\
    & ...,s_R(n),...,s_R(n-(L+L')+1)]^T
\end{aligned}
\end{align}
\noindent where $L'$ is chosen such that $ML'\geq R(L+L')$. Notice that $M$ has to be greater than $R$, and both matrices $\mathbf{Y}$ and $\mathbf{S}$ consists of Hankel blocks. 

The mixing matrix $\mathbf{H}$ is equal to
\begin{equation}
    \label{eq:H}
    \mathbf{H}=[\mathbf{H}_1 \quad \dots \quad \mathbf{H}_R]=
    \begin{bmatrix}
            \mathbf{H_{11}} & \dots & \mathbf{H_{1R}}\\
            \vdots & \ddots & \vdots \\
            \mathbf{H_{M1}} & \dots &  \mathbf{H_{MR}}
    \end{bmatrix}
\end{equation}
\noindent whose any block-entry $\mathbf{H}_{mr}$ is the Toeplitz matrix of $h_{mr}(l)$:

\begin{equation}
    \label{eq:H_ij}
    \mathbf{H}_{mr}=
    \begin{bmatrix}
            h_{mr}(0) & \dots & h_{mr}(L) & \dots & 0\\
              & \ddots & \ddots & \ddots &  \\
            0 & \dots & h_{mr}(0) & \dots & h_{mr}(L)
    \end{bmatrix}
    .
\end{equation}

Next, the autocorrelation $\mathbf{R}_{\mathbf{y}}(\tau)$ for a time lag $\tau$ is expressed as:
\begin{align}
    \label{eq:cov}
         \mathbf{R}_{\mathbf{y}}(\tau)&= \mathrm{E}\{\mathbf{y}(n)\mathbf{y}(n+\tau)^T\} = \mathrm{E}\{\mathbf{H}\mathbf{s}(n)\mathbf{s}(n+\tau)^T\mathbf{H}^T\} \nonumber \\
        &=\mathbf{H} \mathbf{R}_{\mathbf{s}}(\tau)\mathbf{H}^T,  \; \; \; \; \forall\tau.
\end{align}

Assuming that the sources are uncorrelated, the matrices $\mathbf{R}_\mathbf{s}(\tau)$ are block-diagonal, i.e. non-block-diagonal terms representing the correlations between different sources are 0. Therefore, the output autocorrelation matrix $\mathbf{R}_\mathbf{y}(\tau)$ is written as the block-diagonal matrix $\mathbf{R}_\mathbf{s}(\tau)$ multiplied by the mixing matrix $\mathbf{H}$ from the left and by $\mathbf{H}^\text{T}$ from the right. Then, stacking the set of output autocorrelation matrices $\mathbf{R}_\mathbf{y}(\tau)$ for various $\tau$ values will give rise to a tensor $\boldsymbol{\mathcal{T}}$ that admits a so-called block-term decomposition (BTD). More specifically, $\boldsymbol{\mathcal{T}}$ can be written as a sum of low-multilinear rank tensors, in this specific case a rank of $(L+L',L+L',\cdot)$ \citep{b24}. Due to the Hankel-block structure of $\mathbf{Y}$ and $\mathbf{S}$, $\mathbf{R}_{\mathbf{y}}(\tau)$ and $ \mathbf{R}_{\mathbf{s}}(\tau)$ are Toeplitz-block matrices. Note that the number of time-lags to be included is a hyperparameter of the algorithm, and we take it as equal to the filter length in this work.

\begin{figure}[H]
    \centering
    \includegraphics[width=.7\textwidth]{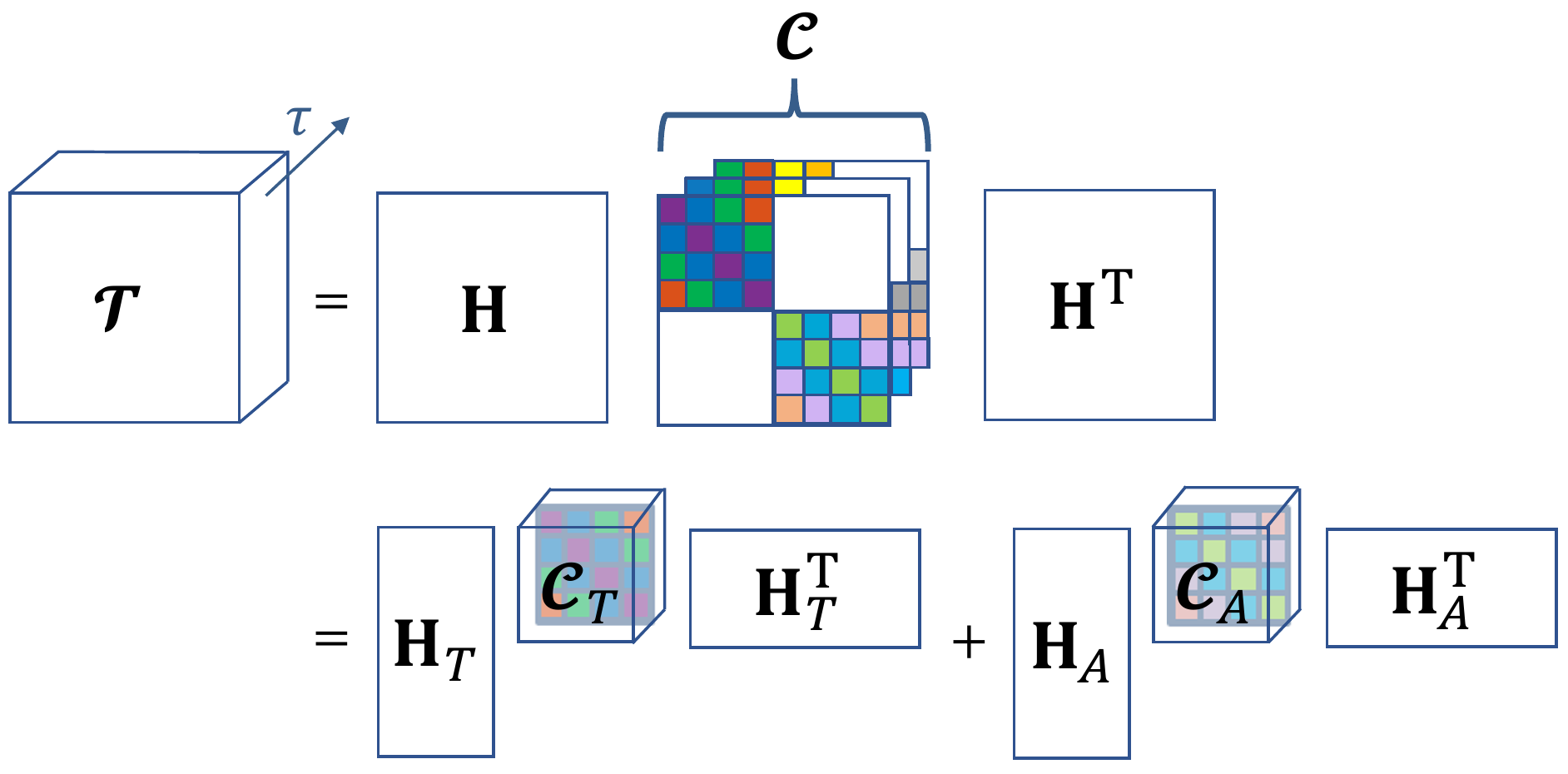}
    \caption{A demonstration of BTD for $R=2$. The tensor $\boldsymbol{\mathcal{T}}$ of stacked measurement autocorrelations $\mathbf{R}_\mathbf{y}(\tau)$, $\forall \tau$ is first expressed in terms of the convolutive mixing matrix $\mathbf{H}$ and a core tensor $\boldsymbol{\mathcal{C}}$ which shows the stacked source autocorrelations $\mathbf{R}_\mathbf{s}(\tau)$, $\forall \tau$. Each $\mathbf{R}_\mathbf{s}(\tau)$ corresponds to a frontal slice of $\boldsymbol{\mathcal{C}}$ and exhibits a block-diagonal structure with inner Toeplitz-blocks. Note that, each slice comes as a lagged version of the preceeding slice. $\boldsymbol{\mathcal{T}}$ is decomposed into $R=2$ terms, each of which contains a core tensor ($\boldsymbol{\mathcal{C}}_T$ or $\boldsymbol{\mathcal{C}}_A$, which represents the autocorrelation of the corresponding source) and a block column of $\mathbf{H}$ ($\mathbf{H}_T$ or $\mathbf{H}_A$).}
    \label{fig:btd_im}
\end{figure}

The decomposition for $R=2$ is illustrated in Fig. \ref{fig:btd_im}. Considering our signal model, where we have defined two types of sources, we can rewrite the block-columns of $\mathbf{H}=[\mathbf{H}_1 \; \mathbf{H}_2]$ (Eq. \ref{eq:H}) simply as $\mathbf{H}=[\mathbf{H}_T \; \mathbf{H}_A]$ instead. Here, $\mathbf{H}_T$ relates to the task-source, i.e. includes the region-specific HRFs, whereas $\mathbf{H}_A$ includes the region-specific scalings of the artifact source. 

In addition, we impose a shape constraint to the HRFs such that they are physiologically interpretable. For this purpose, we employed the model described in \citep{b3}, which is a fUS-based adaptation of the well-known canonical model used predominantly in fMRI studies \citep{b7} for depicting CBF or CBV changes, where the second gamma function leading to the undershoot response is removed from the canonical model, resulting in a reduced number of parameters. This model expresses the HRF in terms of a single gamma function defined on a parameter set $\boldsymbol{\theta}$ as below:
\begin{equation}
    \label{eq:gamma}
        f(t,\boldsymbol{\theta}) = \theta_1(\Gamma(\theta_2)^{-1}  \theta_3^{\theta_2}t^{\theta_2-1}\rm{e}^{-\theta_3t})
\end{equation}
\noindent where $\theta_1$ is the scaling parameter to account for the strength of an HRF and the rest of the parameters define the shape of the HRF. % In addition, we apply a range constraint to the parameters such that they are between a lower bound ($\boldsymbol{\beta}_l$) and an upper bound ($\boldsymbol{\beta}_u$), except for the scaling factor $\theta_1$. The range constraints further ensure that the resulting shapes are physiologically meaningful, as well as help with the time-shift ambiguity \citep{shift_amb} in blind deconvolution tasks.

Finally, the BTD is computed by minimizing the cost function:
\begin{align}
    \label{eq:cost}
J(\boldsymbol{\mathcal{C}},\boldsymbol{\theta},\mathbf{a}) = \lVert  \boldsymbol{\mathcal{T}} &- \sum_{r \in T} \boldsymbol{\mathcal{C}}_r \times_1 \mathbf{H}_r(\boldsymbol{\theta}_r)  \times_2 \mathbf{H}_r(\boldsymbol{\theta}_r) \nonumber \\
&-\sum_{r \in A} \boldsymbol{\mathcal{C}}_r \times_1 \mathbf{H}_r(\mathbf{a}_r)  \times_2 \mathbf{H}_r(\mathbf{a}_r)
\rVert^2_F 
%\\
%\text{su} & \text {bject to   } \; \boldsymbol{\beta}_l<\boldsymbol{\theta}_r<\boldsymbol{\beta}_u
\end{align}

\noindent while all $\mathbf{H}_r$'s and $\boldsymbol{\mathcal{C}}_r$'s are structured to have Toeplitz blocks. The operator $||\cdot||_F$ is the Frobenius norm. The BTD is implemented using the structured data fusion (SDF) framework, more specifically using the quasi-Newton algorithm \texttt{sdf\_minf}, offered by Tensorlab \citep{b25}.

\subsection{Identifying a Stable Solution for BTD}

For many matrix and tensor-based factorizations, such as the BTD described above, the objective functions are non-convex. As such, the algorithm selected for solving the non-convex optimization might converge to local optimas of the problem \citep{tdunique}. In order to identify a stable solution, it is common practice to run the optimization multiple times, with a different initialization at each run. Finally, a choice needs to be made amongst different repetitions of the decomposition. 

For our problem, each BTD repetition produces $M$ HRFs, characterized by their parameters $\boldsymbol{\theta}_m, m=1,2,\dots,M$. We follow a similar approach as described in \citep{simonstable}, i.e. we cluster the solutions using the peak latencies of the estimated HRFs as features, and aim at finding the most coherent cluster. The steps of our clustering approach are as follows:

\begin{enumerate}
    \item Run BTD $20$ times with random initializations, and from each run, store the following:
    \begin{itemize}
        \item Final value of the cost (i.e., objective) function
        \item $M$ HRFs
    \end{itemize}
    \item Eliminate the $P$ outlier BTD repetitions having significantly higher cost values (We use Matlab's \texttt{imbinarize} for the elimination which chooses an optimal threshold value based on Otsu's method \citep{otsu}, as we expect the best solution to be amongst the low-cost solutions)
    \item Form a matrix with $M$ columns (standing for the peak latencies of $M$ HRFs, these are the features) and $20-P$ rows (standing for the retained BTD repetitions, these are the observations)
    \item Apply agglomerative hieararchical clustering to the columns of the matrix in Step 3
    \item Compute the following intracluster distance metric for each cluster as:
    \begin{equation}
    \label{dist_cluster}
        d_\text{C} = \frac{\max_{c_1,c_2 \in \text{C}} d(c_1,c_2)}{n_\text{C}}
    \end{equation}
    where the numerator gives the Euclidean distance between the two most remote observations inside the cluster $\text{C}$ (known as the complete diameter distance \citep{diameterdist}), and the denominator, $n_\text{C}$, is the number of observations included in $\text{C}$
    \item Determine the most stable cluster as the one having the minimum intracluster distance
    \item Calculate the mean of the estimated HRFs belonging to the cluster identified in Step 6
\end{enumerate}

To sum up, the clustering approach described above assumes that the best possible solution will be low-cost (Step 2), have low intracluster distance (numerator of Eq. \ref{dist_cluster}) and frequently-occurring (denominator of Eq. \ref{dist_cluster}). After we have the final HRF predictions, the last step is to estimate the sources.

\subsection{Estimation of the Source Signals}

The final HRF estimates are reorganized in a Toeplitz-block matrix as shown in Equations \ref{eq:H} and \ref{eq:H_ij}. This gives rise to $\hat{\mathbf{H}}_T$, i. e., the block columns of $\mathbf{H}$ that are of interest. Going back to our initial formulation $\mathbf{Y}=\mathbf{H}\mathbf{S}$, we can estimate the task-related source signals $\mathbf{S}_T$ by: 
\begin{equation}
    \hat{\mathbf{S}}_T=\hat{\mathbf{H}}_T^\dagger \mathbf{Y}
    \label{s_ls}
\end{equation}
where $(.)^\dagger$ shows the Moore-Penrose pseudo-inverse. 

In order to obtain the pseudo-inverse of $\hat{\mathbf{H}}_T$, we used truncated singular value decomposition (SVD). Truncated SVD is a method for calculating the pseudo-inverse of a rank-deficient matrix, which is the case for many signal processing applications on real data, such as for extraction of signals from noisy environments \citep{b26}. We applied this approach by heuristically setting the singular values of $\hat{\mathbf{H}}_T$ to be truncated as the lowest $90\%$ following our simulation study. % To determine the set of singular values to be truncated, we use the optimal threshold derived by \citep{svdthres} in case of a non-square matrix and unknown noise level. We later set the singular values of $\hat{\mathbf{H}}_T$ that are less than this threshold to $0$. 

\section{Experimental Setup and Data Acquisition}

% Female C57BL/6NRj mice (8 weeks old, N = 5 for each modality,
% Janvier Lab, France) were used for in vivo experiments. Animals were
% housed in ventilated, temperature-controlled cages under a 12-hour
% dark/light cycle. Pelleted food and water were provided ad-libitum.
% All in vivo experiments were performed in accordance with the Swiss
% Federal Act on Animal Protection and were approved by the Cantonal
% Veterinary Office Zurich upon the recommendation of the Cantonal
% Commission for Animal Experiments Zurich and are compliant with the
% ARRIVE guidelines.

\begin{figure}[H]
    \centering
    \includegraphics[width=\textwidth]{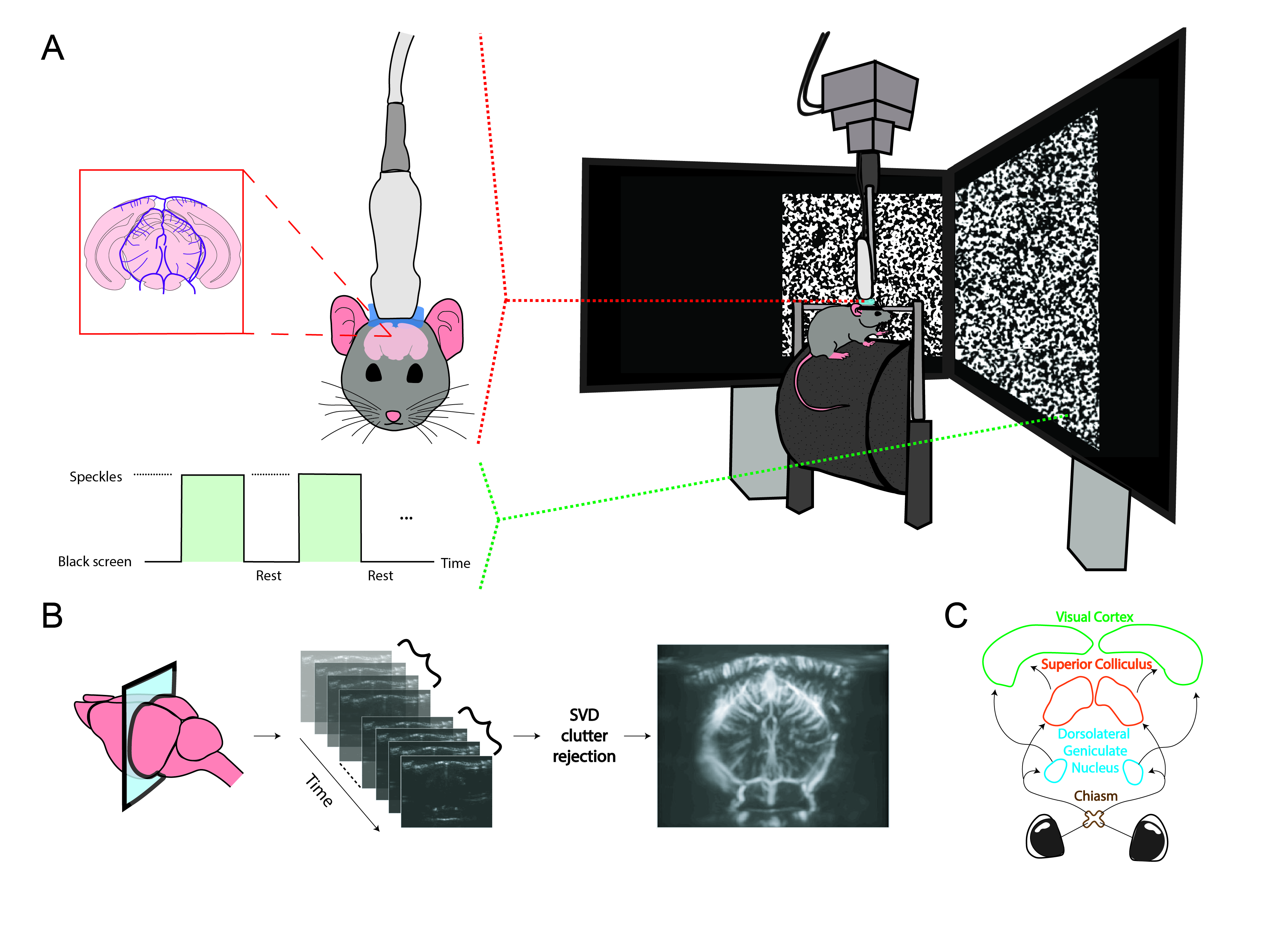}
    \caption{The setup and flowchart for fUS imaging of the ROIs. In A, the experimental setup is shown, with the awake, head-fixed mouse walking on a movable wheel. During an experiment, either a rectangular patch of speckles (stimulus) or an entirely black screen (rest) is displayed across both monitors. In B, the process of forming a PDI is demonstrated for a coronal brain slice. First, back-scattered ultrasonic waves obtained at different imaging angles are beamformed, resulting in compound images. Next, the compound images are progressed to SVD-based clutter filtering in batches in order to remove the tissue motion from the vascular signal. From each filtered batch, a PDI is constructed by computing the power per-pixel. In C, the ROIs that we will focus on in the rest of this paper are shown. The pointed arrows represent the signal flow for processing of visual information.}
    \label{fig:fussetup}
\end{figure}

During our fUS experiment, we displayed visual stimuli to a mouse ($7$-months old, male, C57BL/6J; The Jackson laboratory) while recording the fUS-based HR of its brain via the setup depicted in Fig. \ref{fig:fussetup}. The mouse was housed with food and water \textit{ad libitum}, and was maintained under standard conditions (12/12 h light-darkness cycle, 22℃). Preparation of the mouse involved surgical pedestal placement and craniotomy. %For this, the mouse was put under induction followed by maintenance anaesthesia (5\% and 15 isoflurane in O2 respectively) and was kept at a constant body temperature of 37.6℃ using a heating pad.
First, an in-house developed titanium pedestal (8 mm in width) was placed on the exposed skull using an initial layer of bonding agent (OptiBond™) and dental cement (Charisma\textregistered). Subsequently, a surgical craniotomy was performed to expose the cortex from Bregma -1 mm to -7 mm. %After skull bone removal, the exposed brain was covered with an acoustically transparent, sterilized, 150 μm thick film of Polymethylpentene (TPX™), again with the help of bonding agent and dental cement. 
After skull bone removal and subsequent habituation, the surgically prepared, awake mouse was head-fixed and placed on a movable wheel in front of two stimulation screens (Dell 23,8” S2417DG, 1280 x 720 pixels, 60 Hz) in landscape orientation, positioned at a 45° angle with respect to the antero-posterior axis of the mouse, as well as 20 cm away from the mouse’s eye, similar to \citep{setup}. All experimental procedures were approved \textit{a priori} by an independent animal ethical committee (DEC-Consult, Soest, the Netherlands), and were performed in accordance with the ethical guidelines as required by Dutch law and legislation on animal experimentation, as well as the relevant institutional regulations of Erasmus University Medical Center. 

The visual stimulus consisted of a rectangular patch of randomly generated, high-contrast images - white ``speckles'' against a black background - which succeeded each other with $25$ frames per second, inspired by \citep{param2,param1,speckles}. The rectangular patch spanned across both stimulation screens such that it was centralized in front of the mouse, whereas the screens were kept entirely black during the rest (i.e., non-stimulus) periods. The visual stimulus was presented to the mouse in $20$ blocks of $4$ seconds in duration. Each repetition of the stimulus was followed by a random rest period between $10$ to $15$ seconds. % The high-contrast images were constructed by first creating a custom-made, white square taking up 24° of the visual field, centered against a black background. Every pixel inside this square is assigned a random value using a rotationally symmetric Gaussian low-pass filter of size $10$ and a standard deviation of $3$ as a filtering kernel on the square. These values are then binarized using a threshold at 1.5 to create the aforementioned speckles. 
Before experimental acquisition, a high-resolution anatomical registration scan was made of the exposed brain's microvasculature so as to locate the most ideal imaging location for capturing the ROIs aided by the Allen Mouse Brain Atlas \citep{allen}, and to ensure optimal relative alignment of data across separately performed experiments. Ultimately, during the experiment, functional scans were performed on two slices of the mouse brain; one coronal at Bregma $-3.80$ mm, and one sagittal at Bregma $-2.15$ mm \citep{bregma}.

For data acquisition, $14$ tilted plane waves were transmitted from an ultrasonic transducer (Vermon L$22-14$v, $15$ MHz), which was coupled to the the mouse's cranial window with ultrasound transmission gel (Aquasonic). A compound image was obtained by Fourier-domain beamforming and angular compounding, and non-overlapping ensembles were formed by concatenating $200$ consecutive compound images. We applied SVD based clutter filtering to separate the blood signal from stationary and slow-changing ultrasound signals arising from other brain tissue \citep{svdfilter}. SVD-filtering was performed on each ensemble by setting the first (i.e., largest) $30$\% of the singular values to $0$ and reconstructing the vascular signal of interest from the remaining singular components \citep{fus_setup}. Images were upsampled in the spatial frequency domain to an isotropic resolution of $25\mu$m. % for later registration to the Allen Mouse Brain Atlas \citep{allen}
Finally, a Power-Doppler Image (PDI) was obtained by computing the power of the SVD-filtered signal for each pixel over the ensemble dimension. Hence, the time-series of a pixel (Eq. \ref{eq:convolutive2}) corresponds to the variation of its power across the PDI stream. 

A total of 3 ROIs (SC, LGN and V1) were selected from the captured slices. For this purpose, the data was first parcellated using spatial ICA with $10$ components in both slices \citep{spatial_ica}. A spatial mask was defined based on the spatial signature of the component corresponding to the SC from the coronal; and LGN and V1 from the sagittal slice. To obtain a representative time-series for each ROI, we averaged the time-series of pixels which are captured within the boundaries of each mask. Finally, the ROI time-series were normalized to zero-mean and unit-variance before proceeding with the BTD.

% \textbf{Data Availability}

\section*{Data and Code Availability Statement}

The data and MATLAB scripts that support the findings of this study are publicly available in \href{https://github.com/ayerol/btd_deconv}{https://github.com/ayerol/btd\_deconv}.

\section{Results}

To demonstrate the power of our BTD-based deconvolution approach, the following sections discuss a simulation study and the results of the \textit{in vivo} mouse experiment respectively. In both cases, we consider an EP with repeated stimulation. While we consider a task-related source (expected to be similar to the EP) to affect multiple brain regions through unique HRFs, we also take into account the influence of artifacts and possible hemodynamic changes which are unrelated to the EP on the region HRs. Note that we will use a single additive component (the second term in Eq. \ref{eq:convolutive}) to describe the sources of no interest.
%Note that, in the rest of this work, we will use the artifact-related sources to represent any activity that is not of interest, including the hemodynamic changes induced by neural activity that is unrelated to the EP. 

\subsection{Numerical Simulations} \label{simulation_sec}

We simulated three ROI time-series, each with a unique HRF that is characterized using Eq. \ref{eq:gamma} on a different parameter set $\boldsymbol{\theta}$. We assumed that there are two underlying common sources that build up to the ROI time-series. The first source signal is a binary vector representing the EP. The EP involves $20$ repetitions of a $4$-seconds stimulus (where the vector takes the value $1$) interleaved with $10-15$ seconds of random non-stimulus intervals (where the vector takes the value $0$). This is the same paradigm that will be used later for deconvolution of \textit{in vivo}, mouse-based fUS data (Section \ref{deconv_results}). The EP is assumed to drive the hemodynamic activity in all ROIs, but the measured fUS signals are linked to the EP through possibly different HRFs. The second source signal stands for the artifact component and is generated as a Gaussian process with changing mean, in accordance with the system noise and artifacts modeled in \citep{noisesource}. 

\begin{figure}[H]
    \centering
    \includegraphics[width=\textwidth]{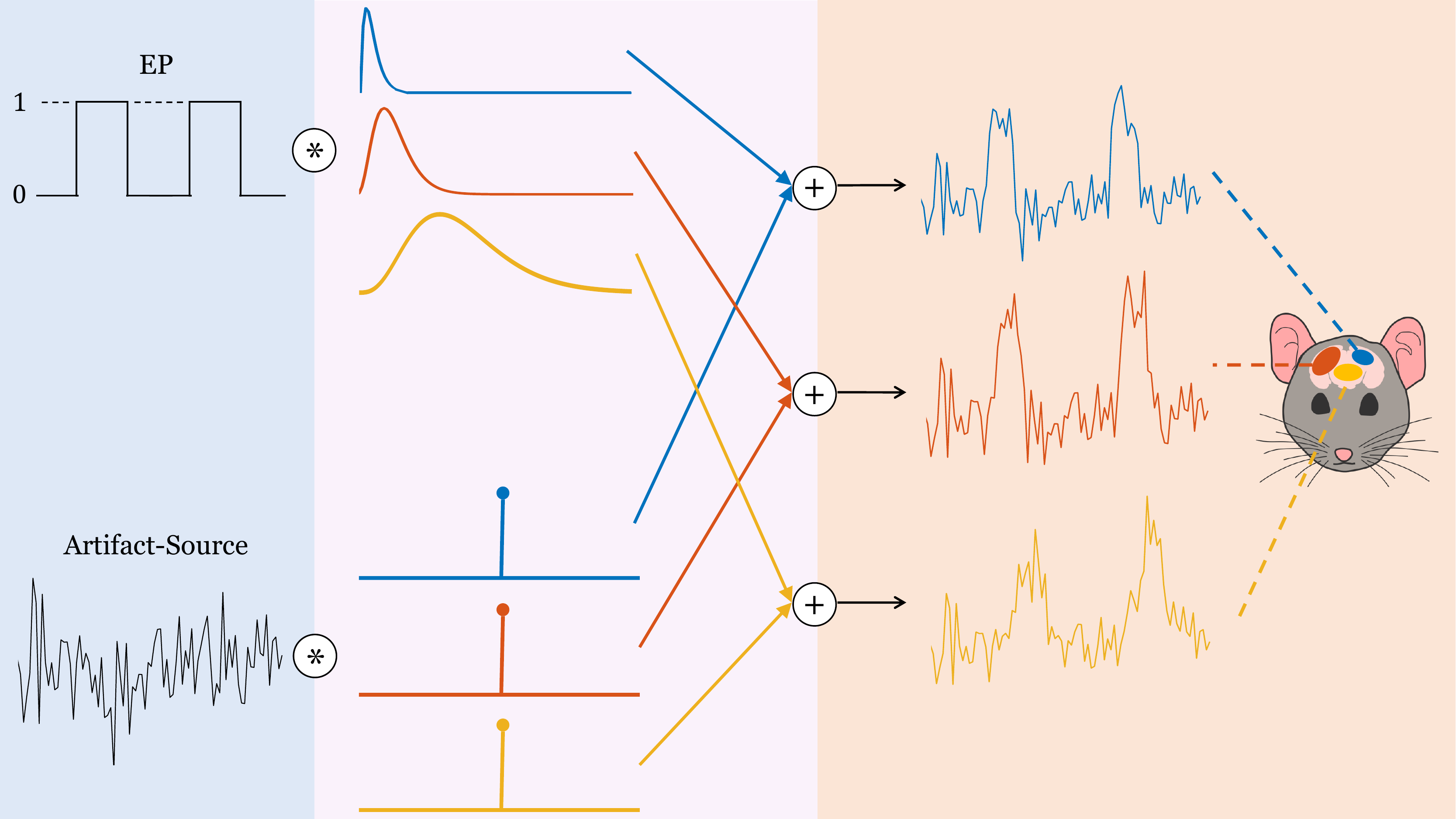}
    \caption{Illustration of the simulator. Both of the simulated sources are shown in the left section, one being task-related (the EP) and one being artifact-related. In the middle section, the convolutive mixing filters are depicted. The filters which are convolved with the EP are the HRFs, whereas the filters which are convolved with the artifact source only differ by their scaling and modeled as impulses, such that their convolution with the artifact source lead to a direct summation on the measured time-series. In the last section, the convolved results are added together to deliver the time-series at each ROI.}
    \label{fig:sim}
\end{figure}

Each ROI time-series is obtained by convolving the corresponding HRF and the common EP, and subsequently adding on the noise source. Note that the variance of the noise source is dependent on the region. In addition, the noise variance values are adjusted in order to assess the performance of the proposed method under various signal-to-noise ratios (SNRs). The data generation steps are illustrated in Fig. \ref{fig:sim}.

We normalized the time-series to zero-mean and unit-variance before proceeding with the BTD. Due to the fact that the true source, in this case the EP, was generated as a binary vector, we binarized the source signal estimated after BTD as well to allow for a fair comparison. More specifically, we binarized the estimated source signal by applying a global threshold, and evaluated the performance of our source estimation by comparing the true onsets and duration of the EP with the predicted ones. 

We performed a Monte Carlo simulation of $100$ iterations for different SNR values. In each iteration, the HRF parameters were generated randomly in such a way that the peak latency (PL) and width (measured as full-width at half-maximum; FWHM) of the simulated HRFs varied between $[0.25,4.5]$ and $[0.5,4.5]$ seconds respectively. These ranges generously cover the CBV/CBF-based HRF peak latencies (reported as $2.1 \pm 0.3$ s in \citep{fus_npixels}, and between $0.9$ and $2$ seconds in \citep{b3,hrf_rng1,hrf_rng2}) and FWHMs (reported as $2.9 \pm 0.6$ s in \citep{fus_npixels}) observed in previous mouse studies.

We defined the following metrics at each Monte Carlo iteration to validate the performance of the algorithm:
\begin{itemize}
    \item For quantifying the match between the estimated and true EP, we calculated the Intersection-over-Union (IoU) between them at each repetition of the EP. For example, if the true EP takes place between $[3,7]$ seconds but this is estimated as $[3.4,7.5]$ seconds, the IoU value will be: $\sfrac{(7-3.4)}{(7.5-3)}=0.8.$ For an easier interpretation, we converted the unit of the IoU ratio to seconds as follows: Since the ideal estimation should give an exact match of $4$ seconds (which corresponds to an IoU of $1$), we multiplied the IoU ratio by $4$. The IoU of $0.8$ in the example above corresponds to a match of $3.2$ seconds. Finally, we averaged the IoU values of $20$ repetitions of the EP to get one final value. 
    \item We computed the absolute PL difference (in terms of seconds) between the true and estimated HRFs, averaged for $M=3$ ROIs.
\end{itemize}

\begin{figure}[H]
  \centering
%   \hspace*{\fill}
  \begin{subfigure}[t]{.49\textwidth}
      \centering
      \includegraphics[width=.87\textwidth]{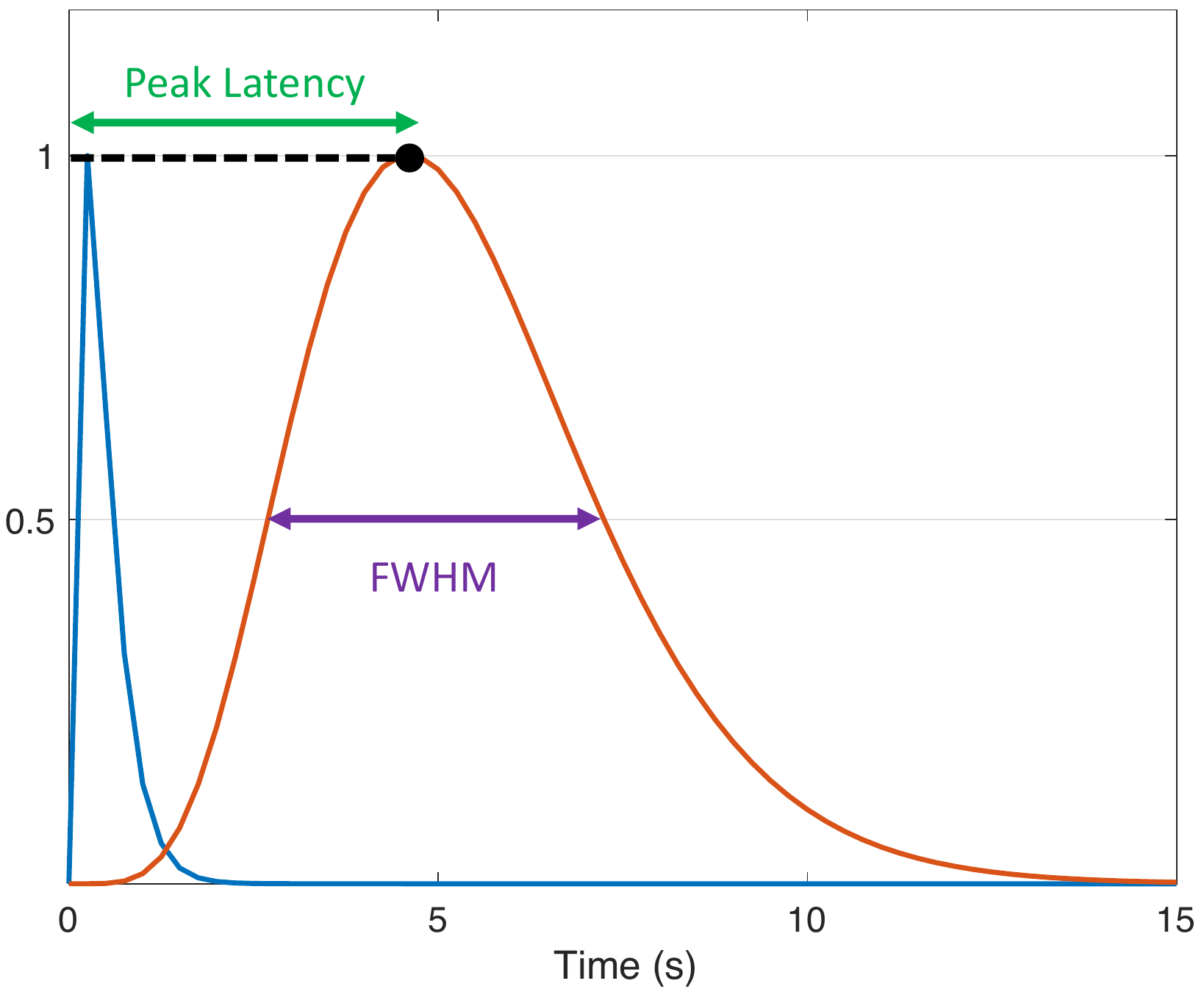}
      \caption{Range of simulated HRFs. The first HRF (blue) has a peak latency and width of $0.25$ and $0.5$ seconds, whereas the second HRF (orange) has both its peak latency and width as $4.5$ seconds respectively. The peak latency and width of the second HRF are also displayed on its plot.}
  \end{subfigure}
  \hspace*{\fill}
  \begin{subfigure}[t]{.47\textwidth}
      \centering
      \includegraphics[width=\textwidth]{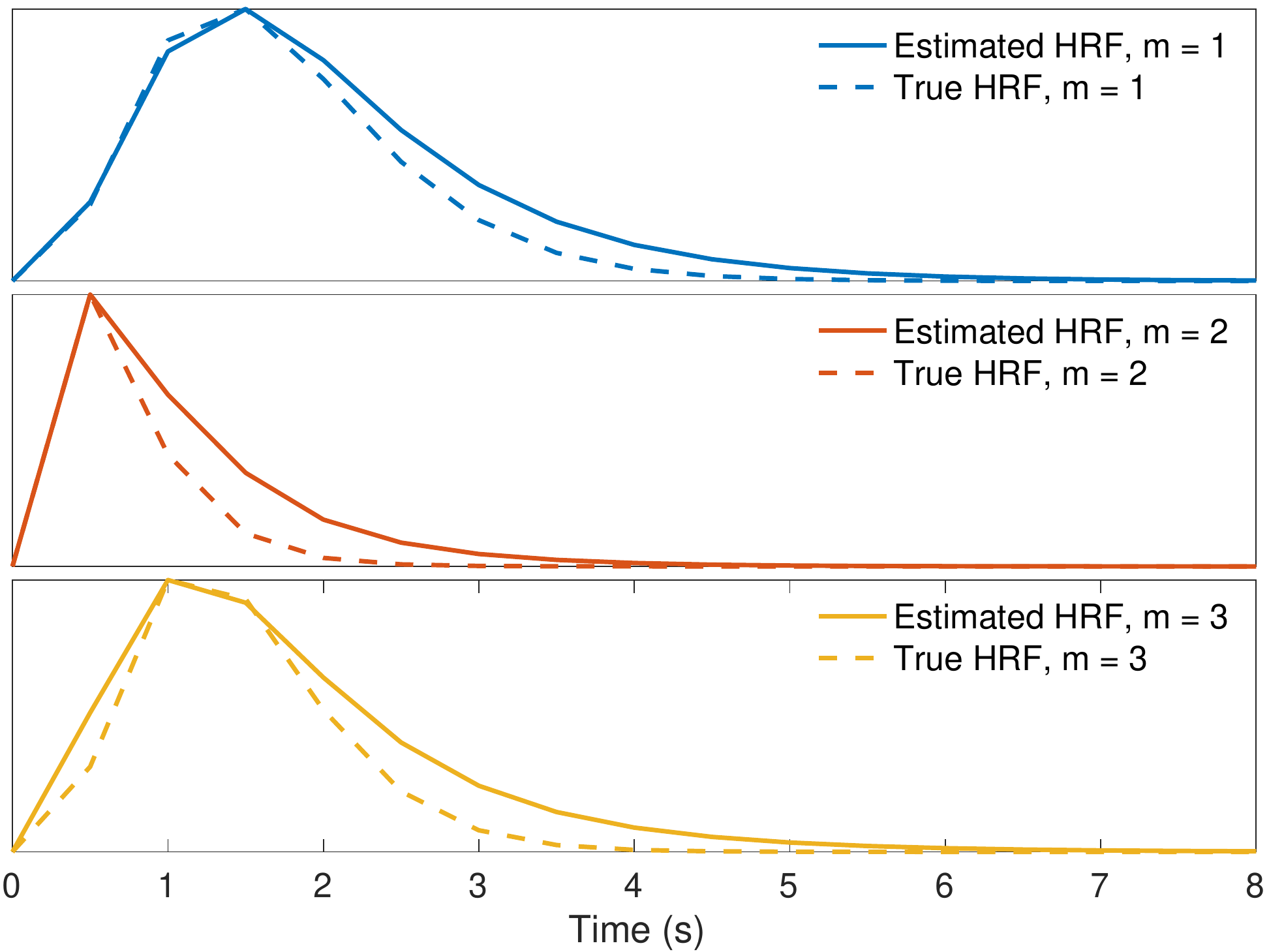}
      \caption{Visualization of the simulated HRFs and their corresponding estimates under $0$ dB SNR (from one Monte-Carlo iteration).}
  \end{subfigure}
  \par\medskip
  \centering
  \begin{subfigure}[t]{.49\textwidth}
      \centering % \unskip\vspace*{-4.57cm}
      \includegraphics[width=\textwidth]{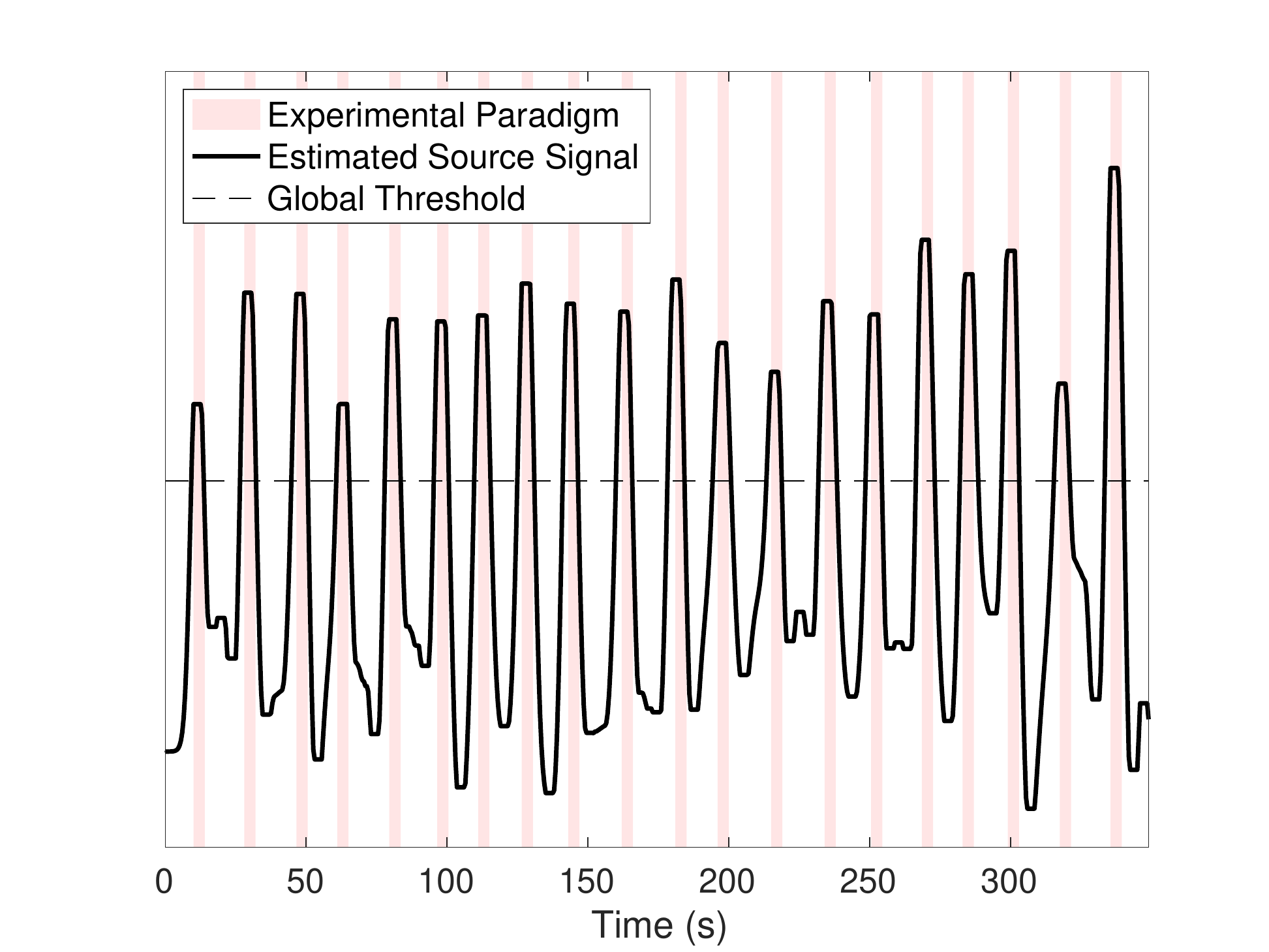} 
      \caption{Visualization of the estimated source signal versus the true EP under $0$ dB SNR (from one Monte-Carlo iteration). For a more precise comparison, we further binarize the estimated source signal by thresholding it as shown.}
  \end{subfigure}
  \hspace*{\fill}
  \begin{subfigure}[t]{.49\textwidth}
      \centering  
      \includegraphics[width=\textwidth]{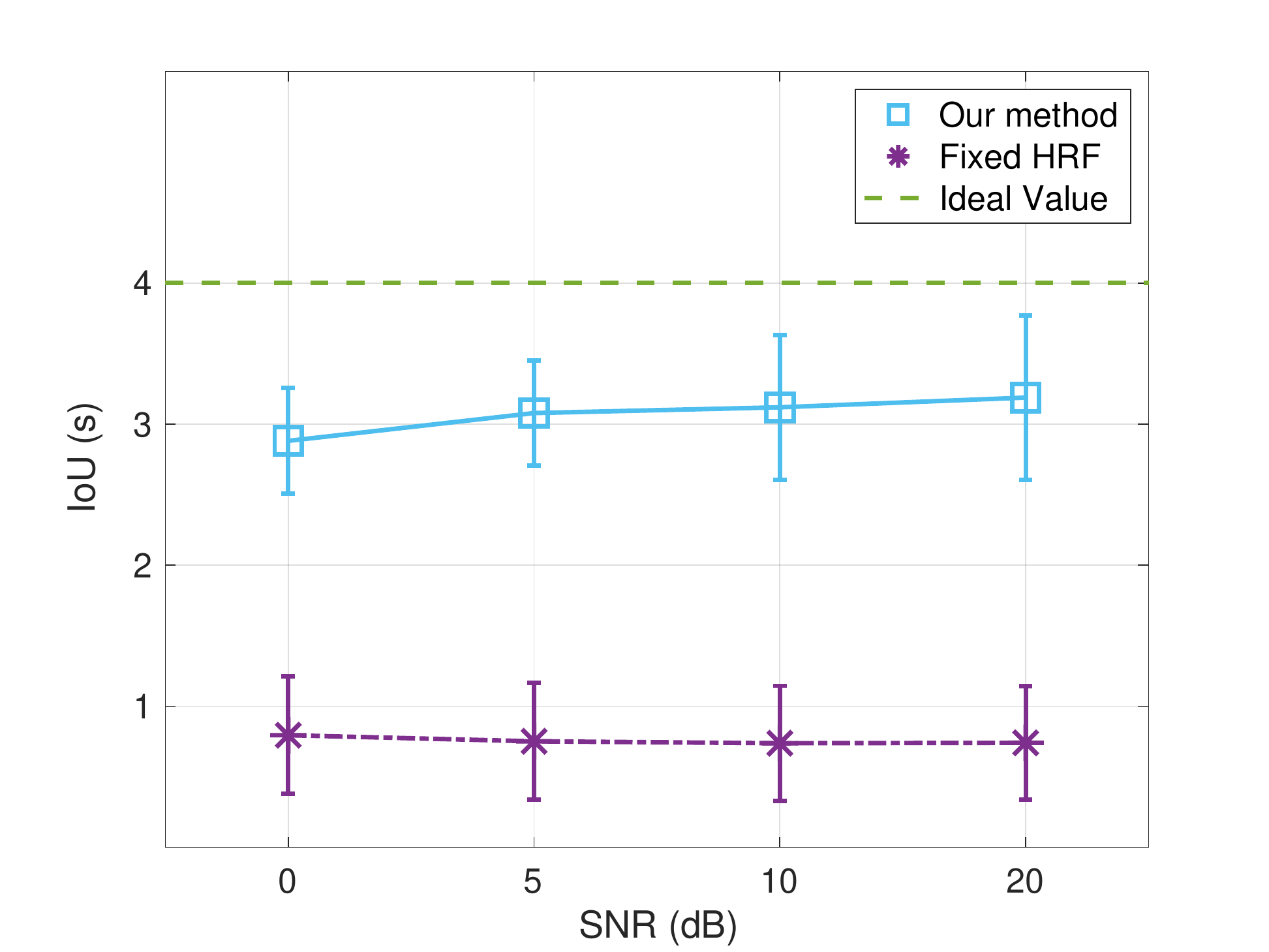}
      \captionsetup{width=.91\linewidth}
      \caption{EP estimation performance with respect to SNR. The markers and errorbars denote the median and standard deviation of the IoU of EP estimation across the Monte-Carlo iterations respectively. When a fixed HRF is assumed, we see that the EP estimation is much less accurate.}
  \end{subfigure}
\caption{Simulation results.}
\label{simresults}
\end{figure}

Simulation results are provided in Fig. \ref{simresults}. Under $0$ dB SNR, the estimated HRFs have an error of $0.3 \pm 0.4$ (median $\pm$ standard deviation) seconds in the peak latencies across the Monte-Carlo iterations. In order to emphasize the importance of incorporating HRF variability in the signal model, we also compared the EP estimation results when a fixed HRF is assumed (the canonical HRF). The results (Fig. \ref{simresults}(d)) show that using a fixed HRF causes a significant decrease in EP estimation performance. In the context of real neuroimaging data, this difference could cause a misinterpretation of the underlying source signals and neurovascular dynamics. 

\subsection{Experimental Data} \label{deconv_results}

\begin{figure}[H]
  \centering
  \begin{subfigure}[t]{.49\textwidth}
      \centering
      \includegraphics[width=\textwidth,trim={2cm 2cm 2cm 1cm},clip]{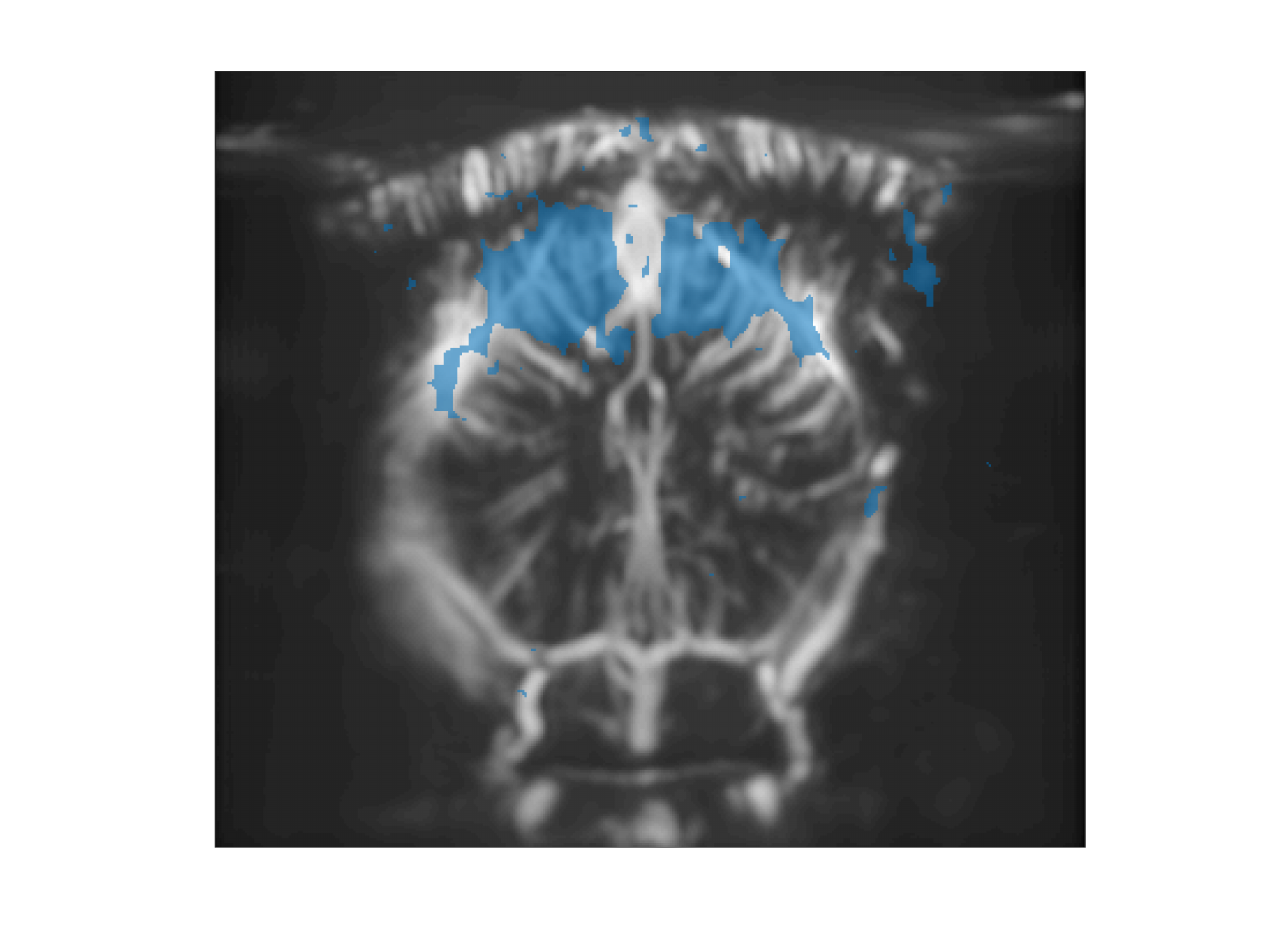}
      \caption{ICA spatial map showing SC (blue).}
  \end{subfigure}
  \hspace*{\fill}
  \begin{subfigure}[t]{.49\textwidth}
      \centering
      \includegraphics[width=\textwidth,trim={2cm 2cm 2cm 1cm},clip]{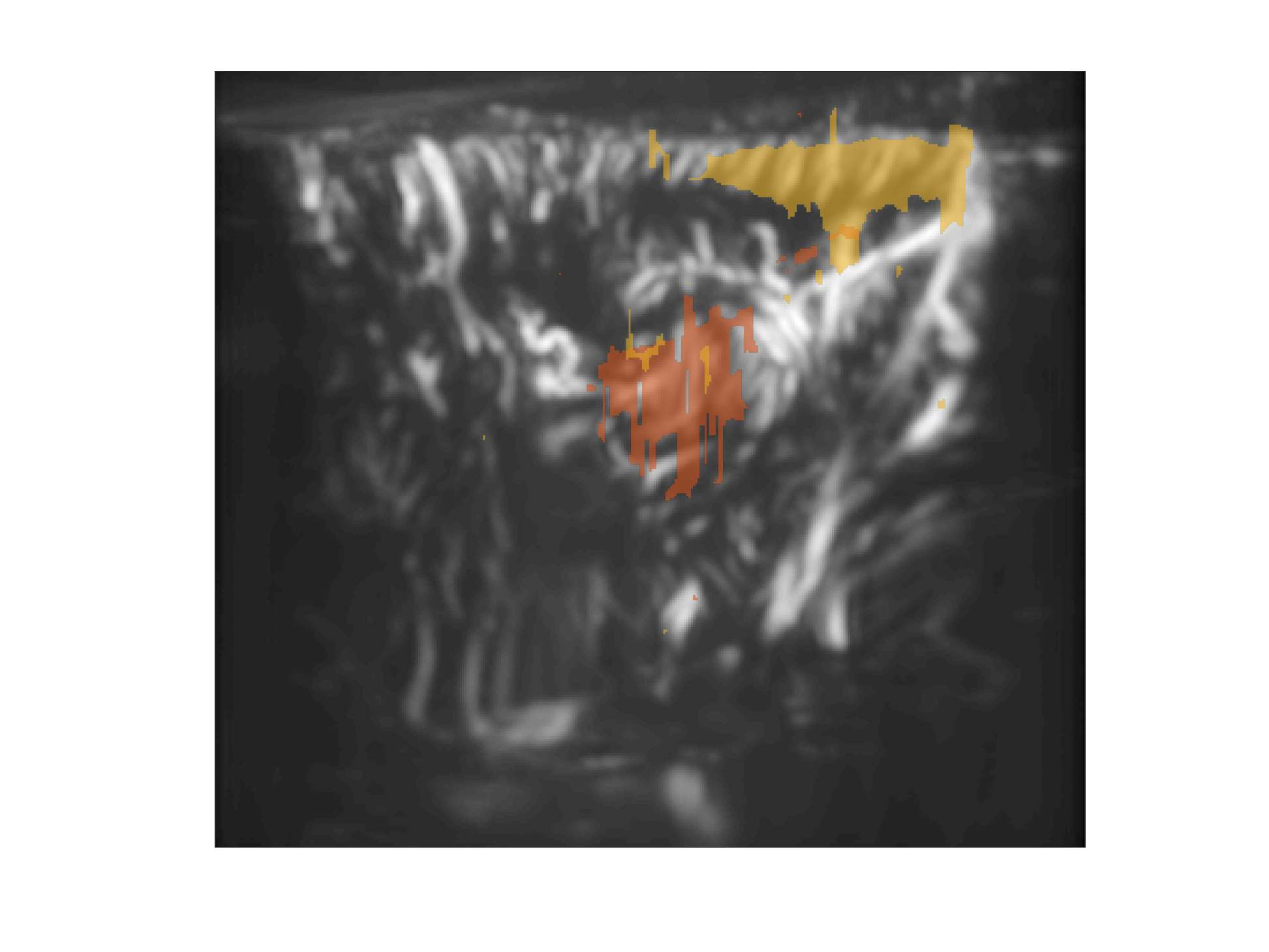}
      \caption{ICA spatial maps showing LGN (orange) and V1 (yellow).}
  \end{subfigure}
  \par\medskip
  \centering
  \begin{subfigure}[t]{.485\textwidth}
      \centering
      \includegraphics[width=\textwidth]{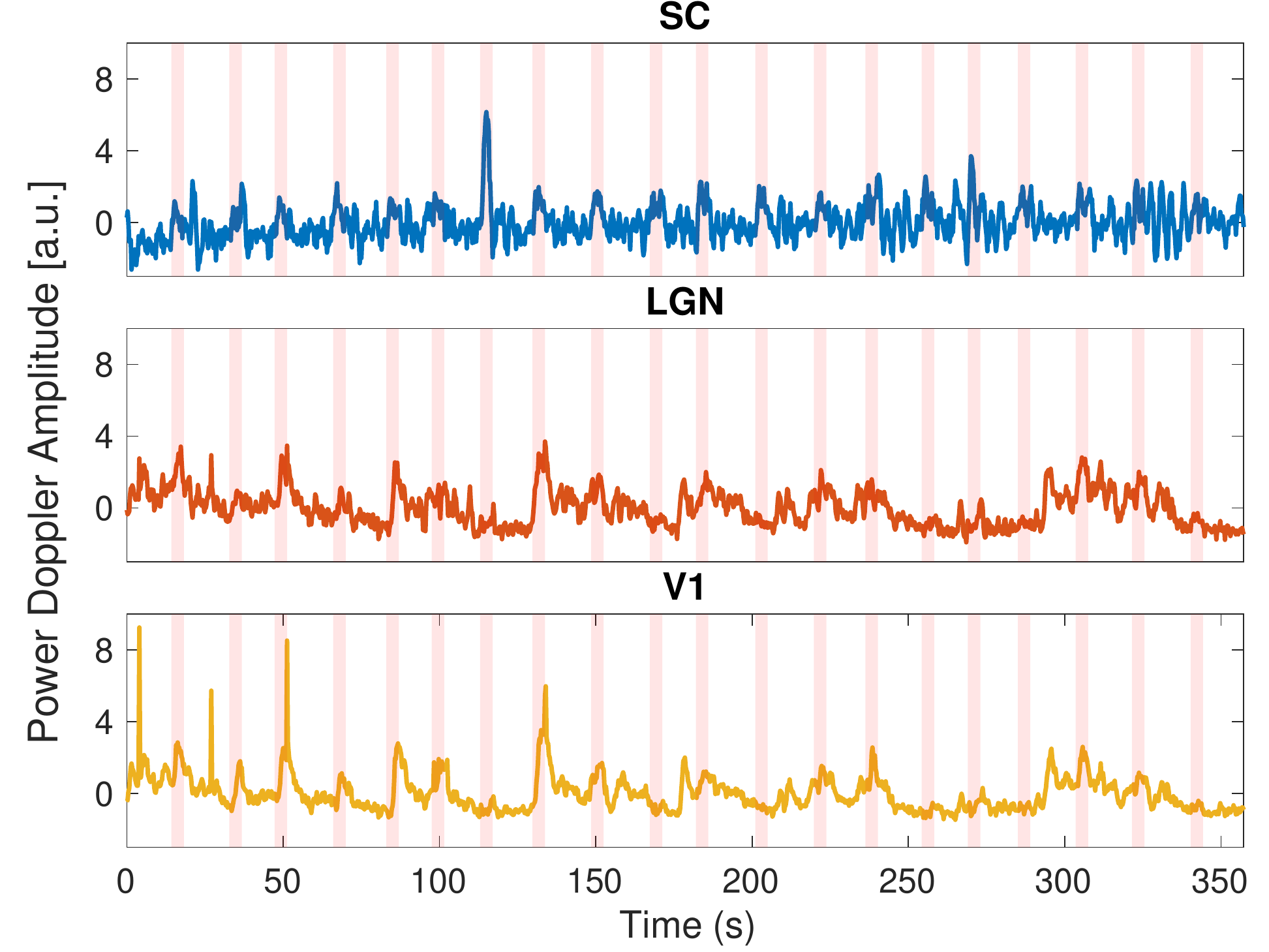}
      \caption{The normalized fUS responses in SC, LGN and V1 (the experimental paradigm is displayed in the background of the plots).}
  \end{subfigure}
  \hspace*{\fill}
  \begin{subfigure}[t]{.495\textwidth}
      \centering
      \includegraphics[width=\textwidth]{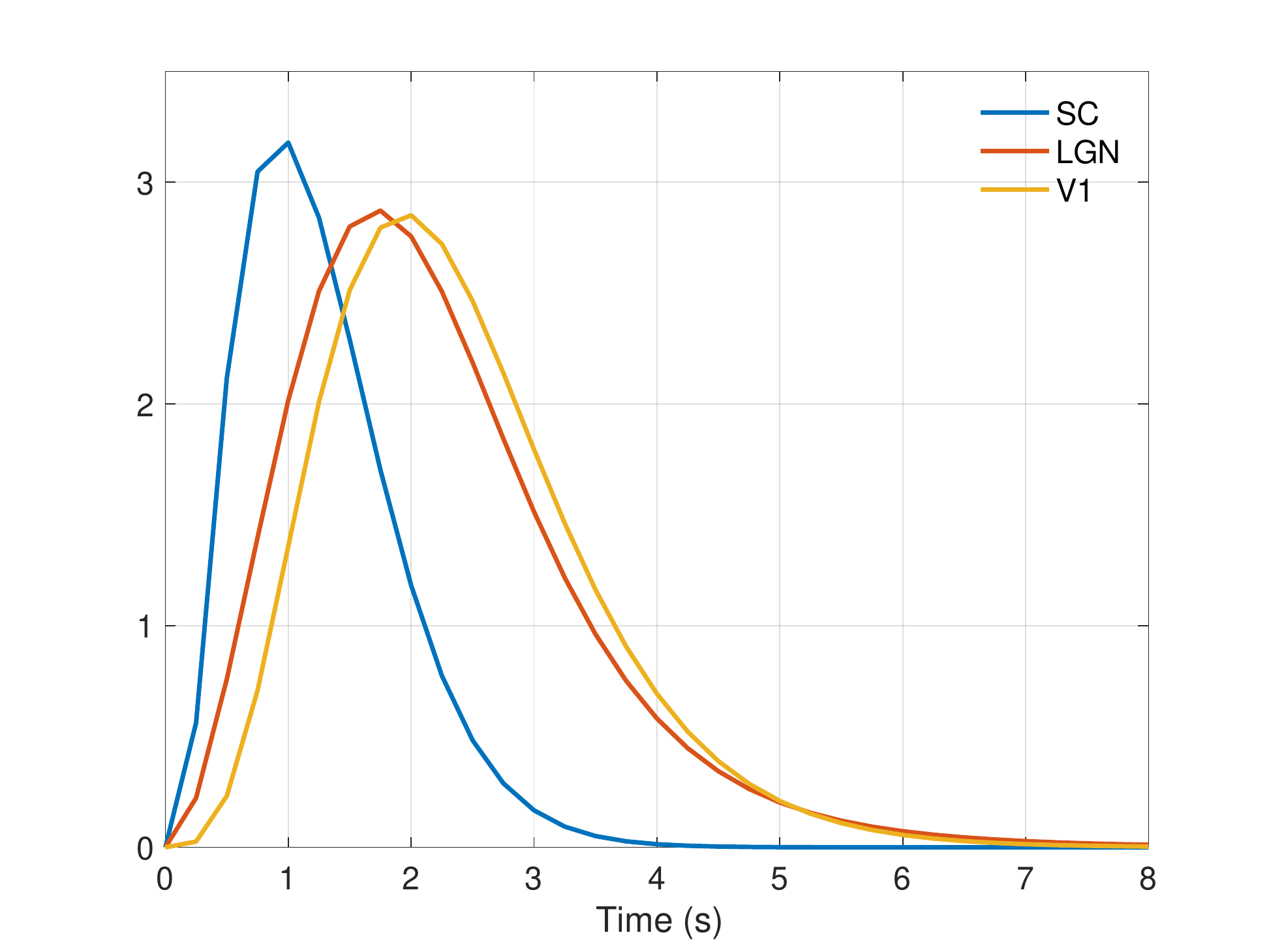}
      \caption{Estimated HRFs.}
  \end{subfigure}
  \par\medskip
  \centering
  \begin{subfigure}[t]{.49\textwidth}
      \centering
      \includegraphics[width=\textwidth]{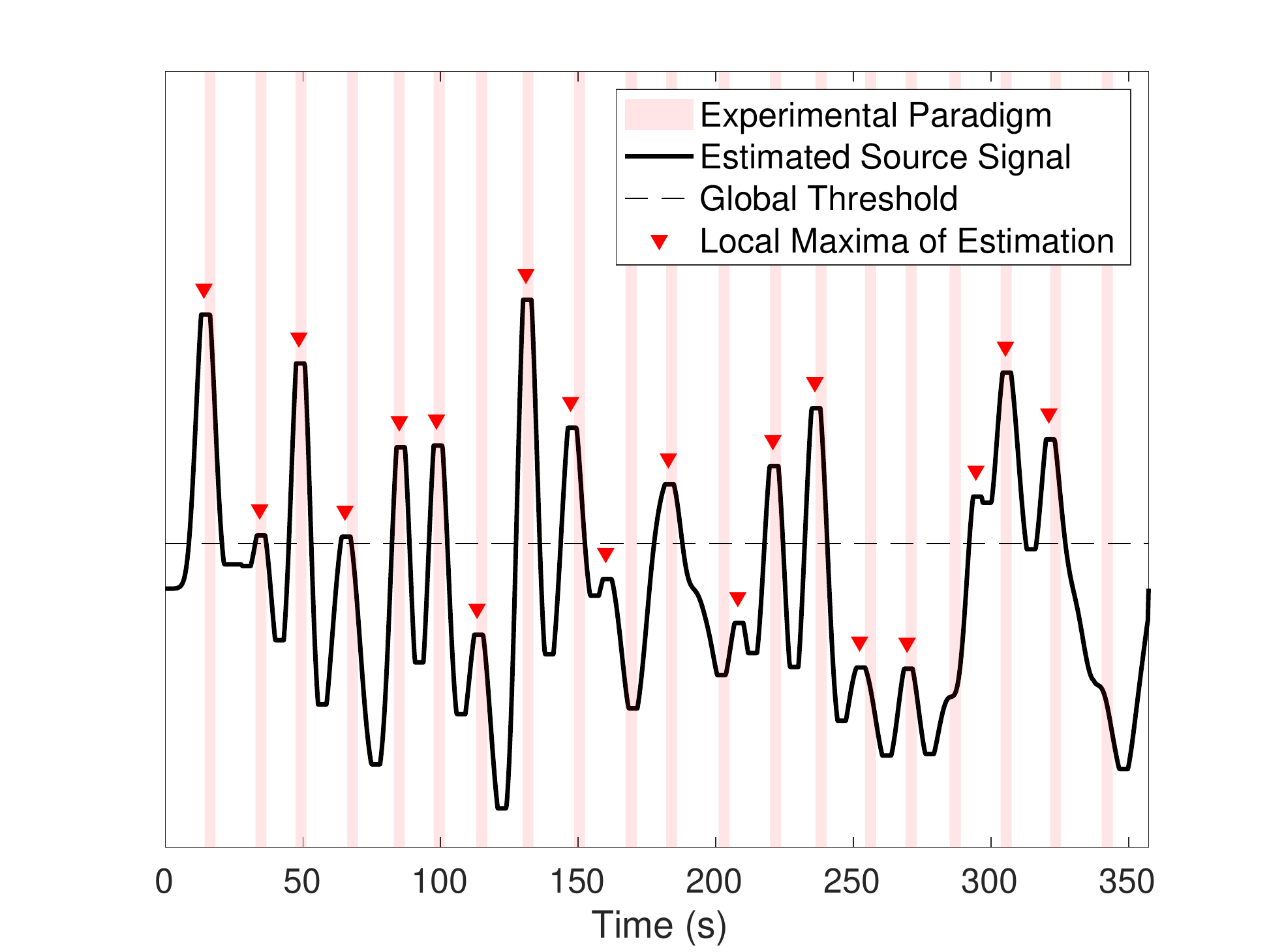}
      \caption{Estimated source signal.}
  \end{subfigure}
  \hspace*{\fill}
  \begin{subfigure}[t]{.49\textwidth}
      \centering
      \includegraphics[width=\textwidth]{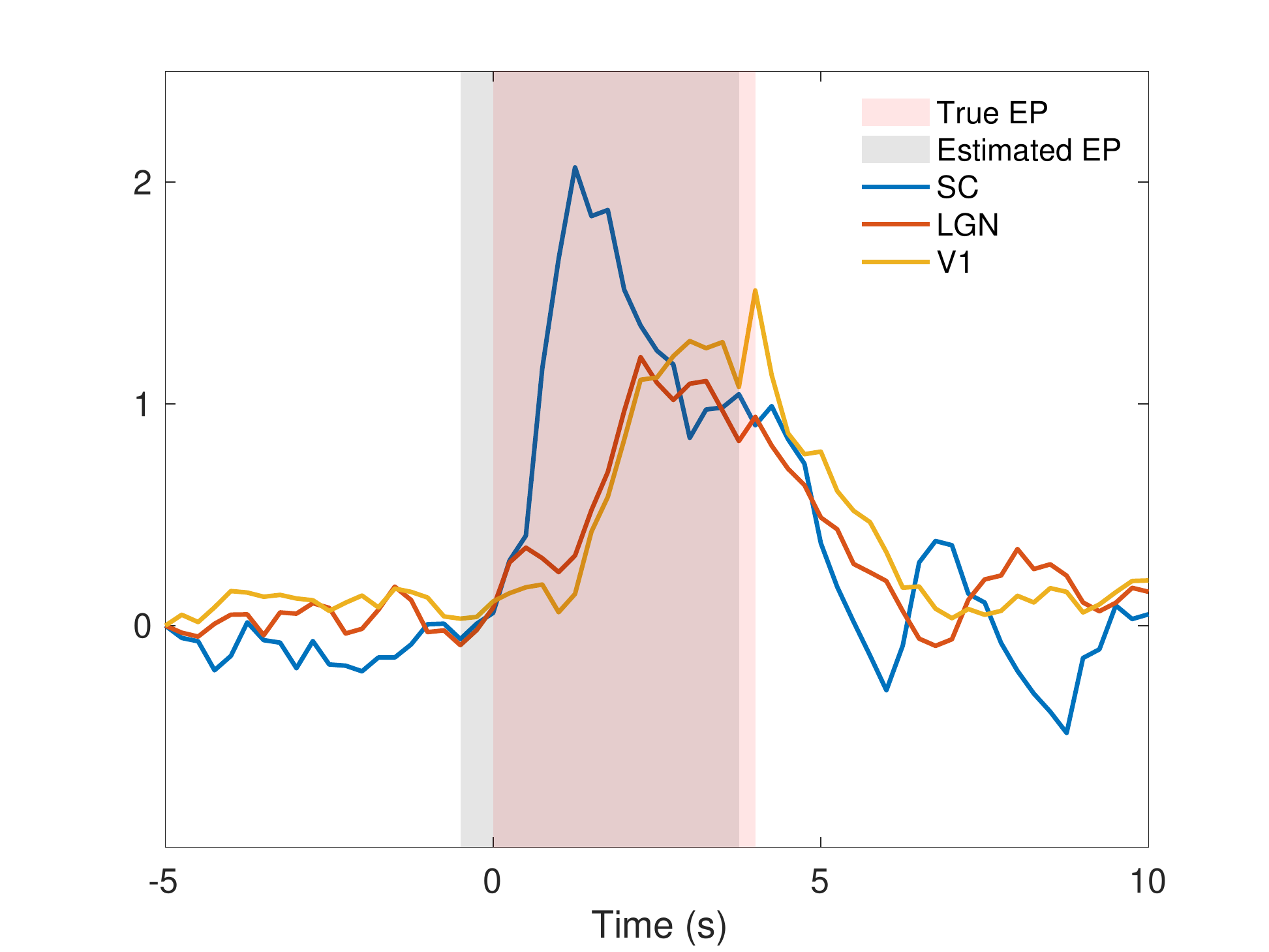}
      \captionsetup{width = .9\textwidth}
        \caption{True EP, estimated EP and normalized responses in SC, LGN and V1, all averaged across stimulus repetitions.}
        \label{reps_regions}
  \end{subfigure}
  \par\medskip
\caption{Deconvolution results on fUS data. Figures (a) and (b) show the ROIs determined with ICA. The regional fUS responses are displayed in (c). HRF and source signal estimation results are given in (d) and (e-f) respectively.}
\label{hrf_exp}
\end{figure}

The selected ROIs are displayed in Fig. \ref{hrf_exp}(a) and Fig. \ref{hrf_exp}(b), showing SC in the former; LGN and V1 in the latter plot. The raw, normalized fUS time-series belonging to each region are displayed in Fig. \ref{hrf_exp}(c). By deconvolving this multivariate time-series data, we estimated the region-specific HRFs and the underlying source signal of interest. In Fig. \ref{hrf_exp}(d), the estimated HRFs are provided. Our results point to a peak latency of $1$ s in SC, $1.75$ s in LGN and $2$ s in V1. Similarly, the FWHMs are found as $1.25$ s in SC, $1.75$ s in LGN and $1.75$ s in V1. These results reveal that SC gives the fastest reaction to the visual stimulus amongst the ROIs, followed by the LGN. In addition, the HRF in SC is observed to be steeper than in LGN and V1. 

Fig. \ref{hrf_exp}(e) demonstrates the estimated source signal of interest. Unlike the simulations, we see that the source signal exhibits a substantial variation in amplitude. In order to interpret this behavior of the estimated source signal, we further investigated the raw fUS signals shown in Fig. \ref{hrf_exp}(c). When the responses given to consecutive repetitions of the stimulus are compared within each region, it can be observed that SC reacts most consistently to the stimulus, while the reproducibility of the evoked responses in LGN and V1 (particularly in V1) are much lower, especially in the second half of the repetitions. To better quantify and compare the region-specific differences in response-variability, we computed the Fano factor (FF) as the ratio of the variance to mean peak amplitude of each region's post-stimulus response \citep{ffmaxamp}, defined in a window $[0,10]$ seconds after a stimulus has been shown. We found an FF value of $0.23, 0.42$ and $0.8$ respectively for SC, LGN and V1. These findings indicate that the consistency of the HR strength is halved from SC to LGN, and again from LGN to V1. 

We can even see cases where there is almost no reaction (as detected by fUS) to the stimulus in V1, such as in repetitions $10, 12, 15, 16$ and $20$. These repetitions coincide with the points in Fig. \ref{hrf_exp}(e) wherein the most considerable drops in the estimated source signal were observed. As such, the variability of responses can explain the unexpected amplitude shifts of the estimated source signal. 

Due to its changing amplitude, comparing the estimated source signal to the EP becomes a more challenging task than in simulations, as binarization using a single global threshold would not work well (Fig. \ref{hrf_exp}(e)). However, it is still possible to observe local peaks of the estimated source signal occurring around the times that the stimulus was shown. While applying a global threshold can uncover $13$ out of $20$ repetitions, with a detection of local peaks, this number increases to $19$ out of $20$ repetitions. After detecting the peaks, we located the time points where for the first time a significant rise (and drop) was observed before (and after) the peak, leading to the starting (and ending) times of the estimated repetitions. Hence, we obtained an estimation of the EP by constructing a binary vector of all $0$'s with the exception of the time periods in between the predicted starting and ending points. In Fig. \ref{hrf_exp}(f), we compared our EP estimation (averaged across repetitions) with the true EP. We can appreciate that our EP estimation is a slightly shifted ($<0.5$ seconds) version of the true EP. Here, we also displayed the responses in SC, LGN and V1 (averaged across repetitions), from which it can be observed that the estimated HRFs follow the same order as the responses during the \textit{in vivo} experiment, as expected.

% If the input signal was assumed to be directly equal to the EP, such response variability would evidence a dynamic system. This follows from the fact that the conventional EP definition assumes that the input signal is exactly the same (equal to $1$) at each repetition of the stimulus, yet the outputs given to the same input are observed to be different at different times. This time-varying behavior would strictly violate the LTI system assumption. However, for our case, we estimate the underlying source signal, which shows explicable amplitude shifts given the responses, meaning that the input signal itself is not the same for when different responses are observed. Although in this study we demonstrate the power of multivariate LTI modelling when both parts of the convolution are kept adaptive, it is also possible to make use of nonlinear and time-varying approaches \citep{balloon,dcm}. While these methods can be adapted to fUS, they come at the cost of introducing more complexity to the problem and thus its solution. %Indeed, we have observed the least response variability in SC, followed by LGN and V1 in the respective order.

Note that the observed trial-by-trial variability in temporal profile across the measured HRs underlines the importance of estimating the source signal. The conventional definition of the EP strictly assumes that the input of the convolution leading to the neuroimaging data (Eq. \ref{eq:singleconv}) is the same ($=1$) at each repetition of the stimulus. This would mean that the exact same input, shown at different times, outputs different responses, which would evidence a dynamic system \citep{balloon,dcm}. However, estimating the source signal allows for a non-binary and flexible characterization of the input, and thus LTI modelling \emph{can} remain plausible. Although extensive analysis of the repetition-dependent behavior of the vascular signal is beyond the scope of this work, we will discuss its possible foundations in the next section.

\newpage 

\section{Discussion} \label{sec_disc}

In this study, we considered the problem of deconvolving multivariate fUS time-series by assuming that the HRFs are parametric and source signals are uncorrelated. We formulated this problem as a block-term decomposition, which delivers estimates for the source signals and region-specific HRFs. We investigated the fUS response in three ROIs of a mouse subject, namely the SC, LGN and V1, which together compose significant pathways between the eye and the brain. 

The proposed method for deconvolution of the hemodynamic response has the advantage of not requiring the source signal(s) to be specified. As such, it can potentially take into account numerous sources besides the EP, that are unrelated to the intended task and/or outside of the experimenters' control. As mentioned in the Introduction, imaged responses might encompass both stimulus-driven and stimulus-independent signals \citep{Xi2020DiverseCN,stringer}. The experimental subject may experience a state of so-called ``quiet wakefulness'' \citep{quitewakeful}, or ``wakeful rest'' \citep{wakefulrest}; a default brain state, during which the unstimulated and resting brain exhibits spontaneous and non-random activity \citep{Gilbert}. This exemplifies how a large fraction of the brain's response is not only triggered by the EP, but also highly influenced by the brain's top-down modulation. This assumption is further supported by recent fUS-based research on functional connectivity \citep{Osmanski} and the default mode network in mice \citep{Dizeux}. Other types of unintentional triggers could be spontaneous epileptic discharges \citep{bori_ica_ep}. 

Despite the fact that the proposed solution has the premise of identifying multiple sources, the number of sources is limited by the selected number of ROIs. As we chose to focus on three ROIs, we were bound to assume less, i.e. two, underlying sources. Accordingly, we accounted for one of the sources to be task-related (related to the visual paradigm), whereas all other noise and artifact terms were combined to represent the second source. Our signal model assumes that the task-related source gets convolved with region-specific HRFs, whereas the artifact-related source is additive on the measured fUS data. As such, the signal model intrinsically assumes that the HRF in each studied region is driven by one common source signal. In fact, incorporating more ROIs and thus more sources can achieve a more realistic approximation of the vascular signal due to the aforementioned reasons. However, it should be noted that addition of more sources would introduce additional uncertainties: what should be the number of sources, and how do we match a source with a certain activity? For instance, several sources can represent the spontaneous (or resting state) brain activity, several sources can represent the external stimuli (depending on the number of different types of stimuli used during the experiment), and the remaining sources can represent the noise and artifacts. To model such cases, the simulations should be extended to include more sources. In addition, thorough studies are needed to explore accurate matching of estimated sources to the activity they symbolize. The assignment of sources can indeed require a priori knowledge of the activities, such as expecting a certain activity to be prominent over the others \citep{cpd_sources}, or defining frequency bands for dividing the signal subspace \citep{reswater}.

%Therefore, instead of relying on the EP for characterization of a whole hemodynamic activity, estimating the sources will result in more accurate modelling, and bring more insight into aforementioned spontaneous triggers of interest. 

When we applied our method to \textit{in vivo} mouse-based fUS data, we observed unforeseen amplitude variations in the estimated source signal. To examine this further, we investigated the hemodynamic responses in the selected ROIs across repetitions. We noticed that the response variability in the visual system increases from the subcortical to the cortical level. Consistent with our findings, electrophysiological studies such as \citep{catlgn} report an increase in trial-by-trial variability from subcortex to cortex, doubling from retina to LGN and again from LGN to visual cortex. Variability in responses could be related to external stimulation other than the EP, such as unintended auditory stimulation from experimental surroundings \citep{ito}. % Such auditory 'noise' can affect the imaged responses in areas such as the SC, which not only receives visual information in its superficial layers, but integrates this with the auditory information it receives in its deeper layers [ito]. The variability across trials has also been attributed to coding of non-stimulus information, i.e. movement or behavior [ref1, ref 2, \citep{mouseeye, eyemov}]. As an example, 
In addition, literature points to eye movements as a source of high response variability in V1, a behavior which can be found in head-fixated, but awake mice following attempted head rotation \citep{mouseeye}, which can extraordinarily alter stimulus-evoked responses \citep{eyemov}. % Neuroanatomical data shows that the mouse V1 is highly interconnected with more than just the subcortical regions of LGN and SC, but also neocortical regions, the effects of which could serve as another possible source for variability [REF]. The stimulus itself could have also played a role in the variable response across different regions. V1 responses could have been less variable, had the stimulus been a natural scene, i.e. with broadband spatiotemporal characteristics, rather than simple stimuli, for example (ref, ref).

% If the input signal was assumed to be directly equal to the EP, such response variability would evidence a dynamic system. This follows from the fact that the conventional EP definition assumes that the input signal is exactly the same (equal to $1$) at each repetition of the stimulus, yet the outputs given to the same input are observed to be different at different times. This time-varying behavior would strictly violate the LTI system assumption. However, for our case, we estimate the underlying source signal, which shows explicable amplitude shifts given the responses, meaning that the input signal itself is not the same for when different responses are observed. Although in this study we demonstrate the power of multivariate LTI modelling when both parts of the convolution are kept adaptive, it is also possible to make use of nonlinear and time-varying approaches \citep{balloon,dcm}. While these methods can be adapted to fUS, they come at the cost of introducing more complexity to the problem and thus its solution. %Indeed, we have observed the least response variability in SC, followed by LGN and V1 in the respective order.

%\citep{lgnv1} also reports that the response variability in subcortical areas is significantly lower than in cortical structures, arguing that a source of response variability in V1 might be originating within the transformation between LGN and V1. 

We noted that the SC has the fastest reaction to stimuli, followed respectively by the LGN and V1. As V1 does not receive direct input from the retina, but via LGN and SC, its delayed HRF is consistent with the underlying subcortical-cortical connections of the visual processing pathway, as has also been reported by \citep{brunner,rats,lewis}. What's more, the SC's particular aptness to swiftly respond to the visual stimulus aligns with its biological function to indicate potential threats (such as flashing, moving or looming spots \citep{Gale, Wang, Inayat}).

Compared to our previous BTD-based deconvolution, we have made several improvements in this work. To start with, the current method exploits all the structures in the decomposition scheme. For example, previously the core tensor representing the lagged source autocorrelations was structured to be having Toeplitz slices, yet, these slices were not enforced to be shifted versions of each other. Incorporating such theoretically-supported structures significantly reduced the computation time of BTD by lowering the number of parameters to be estimated. In addition, we increased the robustness of our algorithm by applying a clustering-based selection of the final HRF estimates amongst multiple randomly-initialized BTD runs. Nevertheless, the formulated optimization problem is highly non-convex with many local minima, and the simulations show that there is still room for improvement. For instance, the selection of hyperparameters - namely the HRF filter length, number of time lags in the autocorrelation tensor, and the number of BTD repetitions - affect the performance of the algorithm. In addition, the selection of the ``best'' solution amongst several repetitions of such a non-convex factorization can be made in various ways, such as with different clustering criteria \citep{icasso}. Although many methods have been proposed to estimate the HRF so far, it is challenging to completely rely on one. First and foremost, we do not know the ground truth HRFs within the brain. As such, it is a difficult research problem on its own to assess the accuracy of a real HRF estimate. Furthermore, all methods make different assumptions on the data to perform deconvolution, such as uncorrelatedness of source signals (this study), the spectral characteristics of neural activity \citep{b14}, and Gaussian-distributed noise terms \citep{b16}. While making certain assumptions about the data model might be inevitable, it is important to keep an open mind about which assumption would remain valid in practice, and under which conditions. Hence, further experiments can be performed in the future to explore the limits of our assumptions.

\section{Conclusion}

In this paper, we deconvolved the fUS-based hemodynamic response in several regions of interest along the mouse visual pathway. We started with a multivariate model of fUS time-series using convolutive mixtures, which allowed us to define region-specific HRFs and multiple underlying source signals. By assuming that the source signals are uncorrelated, we formulated the blind deconvolution problem into a block-term decomposition of the lagged autocorrelation tensor of fUS measurements. The HRFs estimated in SC, LGN and V1 are consistent with the literature and align with the commonly accepted neuroanatomical, biological, neuroscientific functions and interconnections of said areas, whereas the predicted source signal matches well with the experimental paradigm. Overall, our results show that convolutive mixtures with the accompanying tensor-based solution provides a flexible framework for deconvolution, while revealing a detailed and reliable characterization of hemodynamic responses in functional neuroimaging data.

% \section*{Declaration of interest}
% The authors declare that there is no conflict of interest.

% \section*{Credit authorship contribution statement}
% \textbf{Aybüke Erol}: Conceptualization, Methodology, Software, Writing - original draft, Visualization. \textbf{Chagajeg Soloukey}: Investigation, Writing – original draft, Writing – review \& editing. \textbf{Bastian Generowicz}: Resources. \textbf{Nikki van Dorp}: Investigation. \textbf{Sebastiaan Koekkoek}: Resources.
% \textbf{Pieter Kruizinga}: Conceptualization, Resources, Writing – review \& editing, Supervision. \textbf{Borbála Hunyadi}: Conceptualization, Methodology, Software, Writing - original draft, Writing – review \& editing, Supervision.

\section*{Acknowlegements}
This study was funded by the Synergy Grant of Department of Microelectronics of Delft University of Technology and the Delft Technology Fellowship.

\newpage

\bibliography{mybibfile}

\end{document}